\documentclass[twocolumn]{aastex631}

\newcommand{\lya}{Ly$\alpha$ }
\newcommand{\lyab}{Ly$\alpha$}
\newcommand{\dchisq}{$\Delta \chi^2$}
\renewcommand{\textbf}[1]{#1}

\usepackage{amsmath}
\shorttitle{MADGICS for DESI LAEs}
\shortauthors{Uzsoy et al.}
\graphicspath{{./}{figures/}}

\begin{document}

\title{Bayesian Component Separation for DESI LAE Automated Spectroscopic Redshifts \\ and Photometric Targeting}

\author[0000-0001-9308-0449]{Ana Sof{\'i}a M. Uzsoy}
\affiliation{Center for Astrophysics $|$ Harvard \& Smithsonian, 60 Garden St., Cambridge, MA 02138, USA}

\author[0000-0002-6561-9002]{Andrew K. Saydjari}
\altaffiliation{Hubble Fellow}
\affiliation{Center for Astrophysics $|$ Harvard \& Smithsonian, 60 Garden St., Cambridge, MA 02138, USA}
\affiliation{Department of Physics, Harvard University, 17 Oxford St., Cambridge, MA 02138, USA}
\affiliation{Department of Astrophysical Sciences, Princeton University,
Princeton, NJ 08544 USA}
\author[0000-0002-4928-4003]{Arjun~Dey}
\affiliation{NSF NOIRLab, 950 N. Cherry Ave., Tucson, AZ 85719, USA}

\author[0000-0001-5999-7923]{Anand~Raichoor}
\affiliation{Lawrence Berkeley National Laboratory, 1 Cyclotron Road, Berkeley, CA 94720, USA}

\author[0000-0003-2808-275X]{Douglas P. Finkbeiner}
\affiliation{Center for Astrophysics $|$ Harvard \& Smithsonian, 60 Garden St., Cambridge, MA 02138, USA}
\affiliation{Department of Physics, Harvard University, 17 Oxford St., Cambridge, MA 02138, USA}
\author[0000-0003-1530-8713]{Eric Gawiser}
\affiliation{Department of Physics and Astronomy, Rutgers, the State University of New Jersey, Piscataway, NJ 08854, USA}

\author[0000-0003-3004-9596]{Kyoung-Soo Lee}
\affiliation{Department of Physics and Astronomy, Purdue University, 525 Northwestern Ave., West Lafayette, IN 47906, USA}

\author[0000-0001-6098-7247]{Steven~Ahlen}
\affiliation{Physics Dept., Boston University, 590 Commonwealth Avenue, Boston, MA 02215, USA}

\author[0000-0001-9712-0006]{Davide~Bianchi}
\affiliation{Dipartimento di Fisica ``Aldo Pontremoli'', Universit\`a degli Studi di Milano, Via Celoria 16, I-20133 Milano, Italy}
\affiliation{INAF-Osservatorio Astronomico di Brera, Via Brera 28, 20122 Milano, Italy}

\author{David~Brooks}
\affiliation{Department of Physics \& Astronomy, University College London, Gower Street, London, WC1E 6BT, UK}

\author{Todd~Claybaugh}
\affiliation{Lawrence Berkeley National Laboratory, 1 Cyclotron Road, Berkeley, CA 94720, USA}

\author[0000-0002-2169-0595]{Andrei~Cuceu}
\affiliation{Lawrence Berkeley National Laboratory, 1 Cyclotron Road, Berkeley, CA 94720, USA}

\author[0000-0002-1769-1640]{Axel~de la Macorra}
\affiliation{Instituto de F\'{\i}sica, Universidad Nacional Aut\'{o}noma de M\'{e}xico,  Circuito de la Investigaci\'{o}n Cient\'{\i}fica, Ciudad Universitaria, Cd. de M\'{e}xico  C.~P.~04510,  M\'{e}xico}

\author{Peter~Doel}
\affiliation{Department of Physics \& Astronomy, University College London, Gower Street, London, WC1E 6BT, UK}

\author[0000-0002-3033-7312]{Andreu~Font-Ribera}
\affiliation{Institut de F\'{i}sica d'Altes Energies (IFAE), The Barcelona Institute of Science and Technology, Edifici Cn, Campus UAB, 08193, Bellaterra (Barcelona), Spain}

\author[0000-0002-2890-3725]{Jaime~E.~Forero-Romero}
\affiliation{Departamento de F\'isica, Universidad de los Andes, Cra. 1 No. 18A-10, Edificio Ip, CP 111711, Bogot\'a, Colombia}
\affiliation{Observatorio Astron\'omico, Universidad de los Andes, Cra. 1 No. 18A-10, Edificio H, CP 111711 Bogot\'a, Colombia}

\author{Enrique~Gazta\~{n}aga}
\affiliation{Institut d'Estudis Espacials de Catalunya (IEEC), c/ Esteve Terradas 1, Edifici RDIT, Campus PMT-UPC, 08860 Castelldefels, Spain}
\affiliation{Institute of Cosmology and Gravitation, University of Portsmouth, Dennis Sciama Building, Portsmouth, PO1 3FX, UK}
\affiliation{Institute of Space Sciences, ICE-CSIC, Campus UAB, Carrer de Can Magrans s/n, 08913 Bellaterra, Barcelona, Spain}

\author[0000-0003-3142-233X]{Satya~Gontcho A Gontcho}
\affiliation{Lawrence Berkeley National Laboratory, 1 Cyclotron Road, Berkeley, CA 94720, USA}

\author{Gaston~Gutierrez}
\affiliation{Fermi National Accelerator Laboratory, PO Box 500, Batavia, IL 60510, USA}

\author[0000-0002-6024-466X]{Mustapha~Ishak}
\affiliation{Department of Physics, The University of Texas at Dallas, 800 W. Campbell Rd., Richardson, TX 75080, USA}

\author{Robert~Kehoe}
\affiliation{Department of Physics, Southern Methodist University, 3215 Daniel Avenue, Dallas, TX 75275, USA}

\author[0000-0002-8828-5463]{David~Kirkby}
\affiliation{Department of Physics and Astronomy, University of California, Irvine, 92697, USA}

\author[0000-0001-6356-7424]{Anthony~Kremin}
\affiliation{Lawrence Berkeley National Laboratory, 1 Cyclotron Road, Berkeley, CA 94720, USA}

\author[0000-0003-1838-8528]{Martin~Landriau}
\affiliation{Lawrence Berkeley National Laboratory, 1 Cyclotron Road, Berkeley, CA 94720, USA}

\author[0000-0001-7178-8868]{Laurent~Le~Guillou}
\affiliation{Sorbonne Universit\'{e}, CNRS/IN2P3, Laboratoire de Physique Nucl\'{e}aire et de Hautes Energies (LPNHE), FR-75005 Paris, France}

\author[0000-0002-1125-7384]{Aaron~Meisner}
\affiliation{NSF NOIRLab, 950 N. Cherry Ave., Tucson, AZ 85719, USA}

\author{Ramon~Miquel}
\affiliation{Instituci\'{o} Catalana de Recerca i Estudis Avan\c{c}ats, Passeig de Llu\'{\i}s Companys, 23, 08010 Barcelona, Spain}
\affiliation{Institut de F\'{i}sica d'Altes Energies (IFAE), The Barcelona Institute of Science and Technology, Edifici Cn, Campus UAB, 08193, Bellaterra (Barcelona), Spain}

\author[0000-0002-2733-4559]{John~Moustakas}
\affiliation{Department of Physics and Astronomy, Siena College, 515 Loudon Road, Loudonville, NY 12211, USA}

\author[0000-0003-3188-784X]{Nathalie~Palanque-Delabrouille}
\affiliation{IRFU, CEA, Universit\'{e} Paris-Saclay, F-91191 Gif-sur-Yvette, France}
\affiliation{Lawrence Berkeley National Laboratory, 1 Cyclotron Road, Berkeley, CA 94720, USA}

\author[0000-0001-7145-8674]{Francisco~Prada}
\affiliation{Instituto de Astrof\'{i}sica de Andaluc\'{i}a (CSIC), Glorieta de la Astronom\'{i}a, s/n, E-18008 Granada, Spain}

\author[0000-0001-6979-0125]{Ignasi~P\'erez-R\`afols}
\affiliation{Departament de F\'isica, EEBE, Universitat Polit\`ecnica de Catalunya, c/Eduard Maristany 10, 08930 Barcelona, Spain}

\author{Graziano~Rossi}
\affiliation{Department of Physics and Astronomy, Sejong University, 209 Neungdong-ro, Gwangjin-gu, Seoul 05006, Republic of Korea}

\author[0000-0002-9646-8198]{Eusebio~Sanchez}
\affiliation{CIEMAT, Avenida Complutense 40, E-28040 Madrid, Spain}

\author{David~Schlegel}
\affiliation{Lawrence Berkeley National Laboratory, 1 Cyclotron Road, Berkeley, CA 94720, USA}

\author{Michael~Schubnell}
\affiliation{Department of Physics, University of Michigan, 450 Church Street, Ann Arbor, MI 48109, USA}
\affiliation{University of Michigan, 500 S. State Street, Ann Arbor, MI 48109, USA}

\author[0000-0002-6588-3508]{Hee-Jong~Seo}
\affiliation{Department of Physics \& Astronomy, Ohio University, 139 University Terrace, Athens, OH 45701, USA}

\author{David~Sprayberry}
\affiliation{NSF NOIRLab, 950 N. Cherry Ave., Tucson, AZ 85719, USA}

\author[0000-0003-1704-0781]{Gregory~Tarl\'{e}}
\affiliation{University of Michigan, 500 S. State Street, Ann Arbor, MI 48109, USA}

\author{Benjamin~Alan~Weaver}
\affiliation{NSF NOIRLab, 950 N. Cherry Ave., Tucson, AZ 85719, USA}

\author[0000-0002-6684-3997]{Hu~Zou}
\affiliation{National Astronomical Observatories, Chinese Academy of Sciences, A20 Datun Road, Chaoyang District, Beijing, 100101, P.~R.~China}



\begin{abstract}

Lyman Alpha Emitters (LAEs) are valuable high-redshift cosmological probes traditionally identified using specialized narrow-band photometric surveys. In ground-based spectroscopy, it can be difficult to distinguish the sharp LAE peak from residual sky emission lines using automated methods, leading to misclassified redshifts. We present a Bayesian spectral component separation technique to automatically determine spectroscopic redshifts for LAEs while marginalizing over sky residuals. We use visually inspected spectra of LAEs obtained using the Dark Energy Spectroscopic Instrument (DESI) to create a data-driven prior and can determine redshift by jointly inferring sky residual, LAE, and residual components for each individual spectrum. 
We demonstrate this method on \textbf{881} spectroscopically observed $z = 2-4$ DESI LAE candidate spectra and determine their redshifts with $>$90\% accuracy when validated against visually inspected redshifts. Using the $\Delta \chi^2$ value from our pipeline as a proxy for detection confidence, we then explore potential survey design choices and implications for targeting LAEs with medium-band photometry. This method allows for scalability and accuracy in determining redshifts from DESI spectra, and the results provide recommendations for LAE targeting in anticipation of future high-redshift spectroscopic surveys.
\end{abstract}



\section{Introduction} \label{sec:intro}
Lyman-$\alpha$ emitters (LAEs) are young, star-forming galaxies with significant emission (EW $ > 20$\AA) in the \lya line \citep{Hayes_2015, Ouchi+20}. Their existence was first theorized by \cite{partridge_are_1967}, and since then, modern surveys have inferred the physical properties of LAEs. They are low-metallicity with little dust extinction \citep{acquaviva_spectral_2011, ouchi_subaruxmm-newton_2008}, and have significantly higher star formation rates (SFRs) than expected for galaxies of their size \citep{pucha_ly_2022}.  They also have relatively low stellar masses \citep{gawiser_physical_2006} and are thought to be progenitors of present-day Milky Way-like galaxies \citep{gawiser_ly-emitting_2007, Ouchi+20}.

The \lya line profile provides clues about underlying physical processes in LAEs \citep{kunth_hst_1998}. Studies using theoretical radiative transfer simulations suggest that the line shape correlates with intrinsic properties of the emission, such as \lya continuum leakage and hydrogen column density \citep{verhamme_3d_2006, verhamme_using_2015}. There is a known relationship between the FWHM of the \lya line and a systemic velocity offset, possibly due to the opacity of neutral hydrogen in the LAE's interstellar medium \citep{verhamme_recovering_2018}.

Population studies of LAEs over a large redshift range provide valuable constraints for models of reionization
\citep{ouchi_statistics_2010,  zitrin_lyman_2015,  Ouchi+20}. LAEs are more common at higher redshifts up to around $z \approx 6$ before their number density gradually reduces when approaching $z = 7-8$ \citep{schenker_line-emitting_2014, Ouchi+20, tang_lyalpha_2024}. Similarly, the \lya luminosity function can be used to constrain the neutral hydrogen fraction $x_\textsc{Hi}  = n_\textsc{Hi} / (n_\textsc{Hi} + n_\textsc{Hii}) $, which climbs from zero at $z < 6$ to $x_\textsc{Hi}>0.6$ at $z \approx 8$ \citep{ouchi_statistics_2010, itoh_chorus_2018}.

LAEs can be used as spatial tracers of the early universe \citep{im_testing_2024}. They have a similar bias to low-redshift massive galaxies and thus extend our probes of large-scale structure to higher redshifts \citep{white_clustering_2024}. LAEs in ``blobs" closely trace galactic protoclusters \citep{dey_spectroscopic_2016, huang_evaluating_2022, ramakrishnan2024odinidentifyingprotoclusterscosmic}. Studies of LAE clustering also provide constraints on the masses of their dark matter halos \citep{ouchi_systematic_2018, herrera2025odinclusteringanalysis14000}.

Large-scale observational studies of LAEs have been key to realizing their potential as cosmological probes.  LAEs have been targeted mainly through spectroscopy and narrow-band photometry, and have been observed from the ground and space, including with Keck \citep{Spinrad+1997, Dey+98, Hu+1998, Hu+1999, Stern+2000, Manning+2000,cooper_web_2023}, Very Large Telescope \citep{Kudritzki+2000}, Hubble Space Telescope \citep{Pascarelle+1998}, Dark Energy Camera \citep[DECam;][]{lee_one-hundred-deg2_2023}, Subaru Hyper-Suprime Cam \citep{kawanomoto_hyper_2018, aihara_second_2019, Kikuta+23}, James Webb Space Telescope \citep{kumari_jades_2024}, and most recently, the Dark Energy Spectroscopic Instrument \citep[DESI;][]{dey_DESI_ODIN}.

 The Dark Energy Spectroscopic Instrument (DESI) has collected spectra of millions of galaxies for precise cosmological measurements \citep{desicollaboration2016, DESI2022.KP1.Instr, DESI2024.I.DR1}. While LAEs are not one of the primary targets, they have been observed by DESI through secondary and tertiary targeting programs, providing spectroscopic follow-up for LAE candidates identified from photometric surveys \citep{dey_DESI_ODIN, raichoor_DESI_HSC}. After DESI completes its initial five-year run, the planned follow-up program (DESI-2) will focus on higher-redshift targets and aims to observe $\approx$1.5 million LAEs \citep{schlegel_spectroscopic_2022, ravi_examining_2024}. 
 
 The current DESI spectroscopic data analysis pipeline \citep{redrock} does not automatically identify or determine redshifts for LAEs, which are instead currently done by visual inspection. In anticipation of LAEs becoming a primary target for DESI-2, it is important to have reliable, automated software pipelines to identify and characterize them. This requires the use of principled data methods that maximize interpretability while being computationally efficient. 

Here, we take a component separation approach to LAE identification and automated redshift determination to provide a robust, scalable, and data-driven alternative to visual inspection. Our automated method determines a redshift for each target and provides a \dchisq{} metric that serves as a proxy for confidence in an LAE detection. We discuss the accuracy of redshift determination and the redshift success rates resulting from targeting LAE targets selected from medium-band (MB) photometry to inform survey design choices in anticipation of DESI-2.

An outline of this paper is as follows: in Section~\ref{sec:data} we describe the spectra being used and how the objects were targeted. In Section~\ref{sec:methods} we describe in detail the methods we use, how we generate our priors for each component, and how they are applied to the data. The results of the redshift pipeline on real and simulated data and applications to LAE targeting are presented in Section~\ref{sec:results} and discussed further in Section~\ref{sec:discussion} before we conclude in Section~\ref{sec:conclusion}.

\section{Data} \label{sec:data}
\subsection{DESI}

DESI is an optical multi-object fiber spectrograph located on the Nicholas U. Mayall 4-meter Telescope at the Kitt Peak National Observatory in Arizona \citep{Snowmass2013.Levi, DESI2016a.Science, DESI2022.KP1.Instr}. It uses robotic fiber positioners that allow it to collect 5,000 high-quality spectra simultaneously \citep{DESI2016b.Instr, FocalPlane.Silber.2023, Corrector.Miller.2023}. DESI spectra span $3600-9800$ \AA\ with $R = \lambda / \Delta \lambda$ between 2000 and 5000, and are binned linearly to $0.8$ \AA\ pixels \citep{desicollaboration2016, FiberSystem.Poppett.2024}. For more information about survey validation and the Early Data Release, see \citet{DESI2023a.KP1.SV, DESI2023b.KP1.EDR}; for Data Release 1, see \citet{DESI2024.I.DR1}, and for key cosmological results, see the DESI key papers \citep{ DESI2024.II.KP3, DESI2024.III.KP4, DESI2024.IV.KP6, DESI2024.V.KP5, DESI2024.VI.KP7A, DESI2024.VII.KP7B}.

 The four main galaxy samples targeted by DESI are bright galaxies, luminous red galaxies (LRGs), emission line galaxies (ELGs), and QSOs \citep{myers_target-selection_2023, SurveyOps.Schlafly.2023}. The DESI Collaboration has an extensive spectroscopic data analysis pipeline that includes automated redshift determination for each of these four main targeted object types \citep{Guy+23, redrock, RedrockQSO.Brodzeller.2023}. The current automated redshift determination pipeline, Redrock, uses a Principal Component Analysis (PCA)-based approach and has templates for stars, galaxies, and quasars \citep{Guy+23, redrock}. DESI has also implemented a significant visual inspection (VI) campaign for tens of thousands of galaxy spectra, where collaboration members look at individual spectra and report a redshift estimate, as well as provide written comments about its spectral type \citep{lan_desi_2023}. Currently, the Redrock redshift pipeline does not yield accurate redshifts for the LAE samples, and instead redshift measurement has been done via visual inspection.

 As DESI is a ground-based instrument, the Earth's atmosphere imprints contaminating ``sky lines" into all target spectra \citep{desicollaboration2016, Guy+23}. All observed spectra undergo sky subtraction, where a sky model is fit based on observed sky spectra and then subtracted from the target spectrum \citep{Guy+23}. Although this generally works well, there are often systematic sky line residuals that remain in target spectra \citep{Guy+23}. For LAEs, which have only one narrow, comparatively faint emission line, these residual sky emission lines can cause line confusion and misclassified redshifts. Having correct redshifts is critical for using LAEs as cosmological probes, and an automated method is necessary for analyzing LAEs at scale. 

Similar work in automated redshift finding has been done for Lyman-break galaxies (LBGs), which share many characteristics with LAEs \citep{dayal_ly_2012}. \cite{ruhlmann-kleider_high_2024}  use a combination of a convolutional neural network and custom LBG templates alongside the DESI data analysis pipeline to identify and determine redshifts for LBG candidates. Their method yields $\sim$80\% efficiency in redshift determination compared to visual inspection.

\begin{figure}
    \centering
    \includegraphics[width=\linewidth]{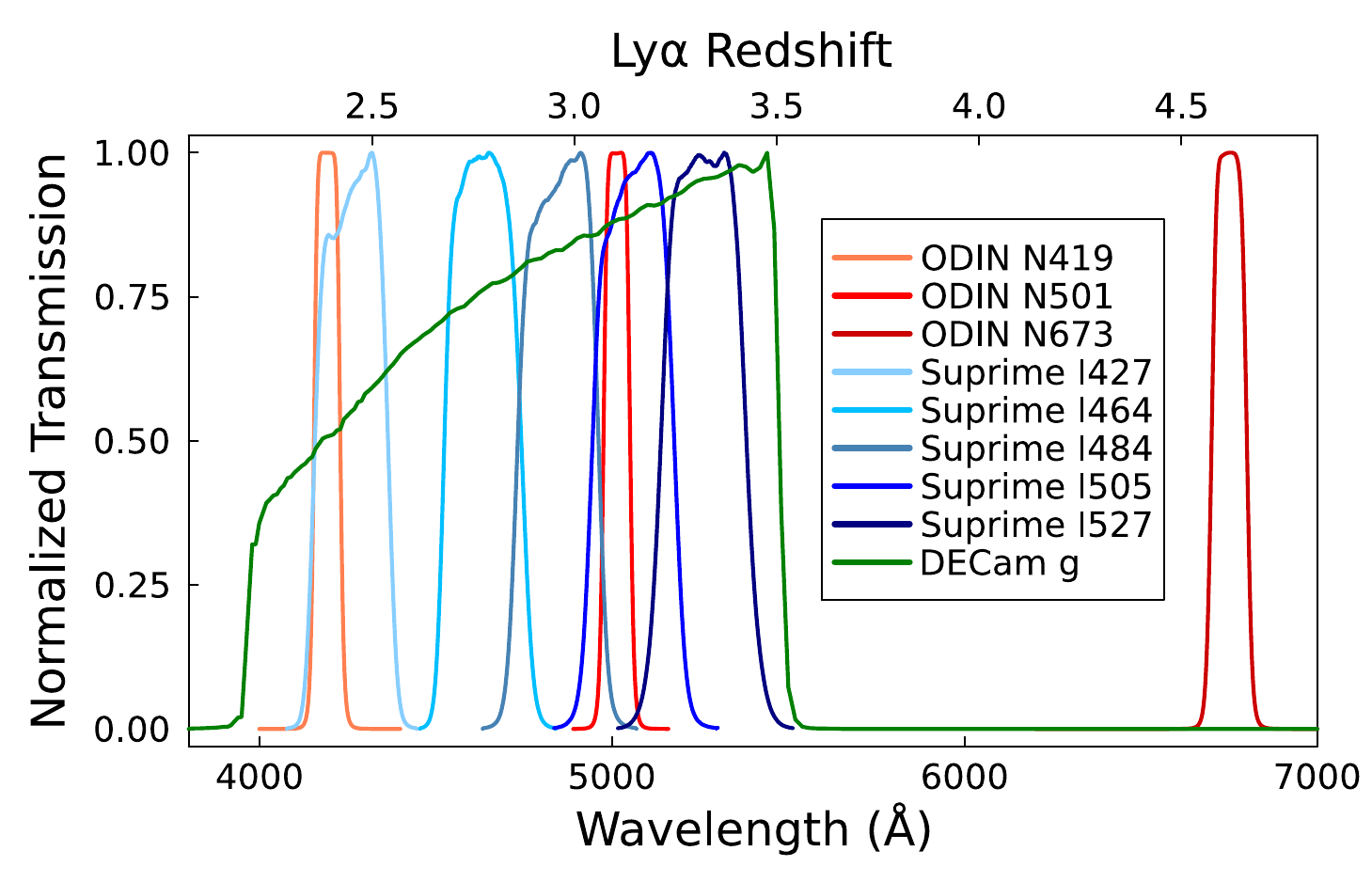}
    \caption{Filter transmission curves (normalized to a maximum transmission of 1) for the N419, N501, and N673 narrow-band filters from ODIN (red curves), I427, I464, I484, I505, and I527 medium-band filters from the Subaru Suprime Cam (blue curves), and the DECam g band (green curve). }
    \label{fig:filters}
\end{figure}

\subsection{The LAE Samples}

For this study, we used two different samples of LAEs targeted by DESI: one selected from the One-hundred deg$^2$ DECam Imaging in Narrowbands (ODIN) survey and the second selected from archival Subaru Suprime-Cam intermediate-band imaging data. 
We treat the narrow-band targeted ODIN LAE candidates, which are a larger sample size, as known LAEs, and use them to create our data-driven LAE prior. This prior is then applied to infer the presence of an LAE and its redshift in spectra of the Subaru LAE candidates, which we use as our test set. The details of the target selection and spectroscopy of these LAE samples are described in more detail elsewhere \citep[see][]{dey_DESI_ODIN, raichoor_DESI_HSC}, but we provide a brief description of these data here.

\subsubsection{ODIN LAEs}
\label{sec:odin_laes}
The One-hundred deg$^2$ DECam Imaging in Narrowbands (ODIN) survey is a photometric survey specifically designed to find potential LAEs by identifying targets with narrow-band excess flux \citep{lee_one-hundred-deg2_2023}. It uses the Dark Energy Camera (DECam) at the prime focus of the 4m Victor Blanco Telescope located at the Cerro Tololo Inter-American Observatory (CTIO) in Chile. ODIN uses three 70--100 \AA\ FWHM narrow-band filters (N419, N501, and N673, seen in Figure~\ref{fig:filters}) to image seven fields covering a total area of $\approx 90~{\rm deg^2}$ to depths of 25.5 -- 25.9 mag \citep{Firestone+23}. \citet{dey_DESI_ODIN} used the ODIN imaging data in the COSMOS and XMM fields in conjunction with archival $grizy$ imaging from the Subaru Hyper-SuprimeCam Subaru Strategic Program survey \citep[]{Aihara+18,aihara_second_2019,Aihara+22} and $grz$ imaging from the DESI Legacy Imaging Surveys \citep[]{Dey+2019} in regions not covered by the HSC imaging. New photometric catalogs of these data were created using Tractor \citep{Tractor2016} and LAE candidates were selected in the COSMOS and XMM fields (see \citet{dey_DESI_ODIN} for details). DESI observations of 11,634 of these candidates were carried out during 2022 and 2023. 

All the spectra were processed by the DESI spectroscopic reduction pipeline, and all spectra were visually inspected by members of the DESI and ODIN collaboration to measure redshifts. The visual inspection results in a redshift estimate (typically determined from the peak of the Ly$\alpha$ emission line) and a quality flag (`VI\_QUALITY') which ranges from 0 to 4. Very briefly, the rubric used classifies quality as: ``0" for no detected flux; ``1" for some weak flux, perhaps continuum, but no redshift estimate; ``2" for a weak feature detected at low signal-to-noise and a tentative redshift; ``3" for a single line redshift (which is reliable when the line is \lyab) by a break, and with no other significant lines in the DESI spectrum; or ``4" for a firm redshift where two or more lines are detected. Each spectrum was inspected by at least two individuals, with the quality flag averaged between them.

The visual inspections also provide labels for the spectra, identifying LAEs, QSOs, and low-redshift galaxy interlopers (see \citet{dey_DESI_ODIN} for further details). QSOs were identified by visual inspectors \textbf{likely} based on the presence of broad emission lines and were noted in the VI comments; \textbf{763} objects \textbf{visually identified as QSOs} are excluded from this sample. From these observations and associated labels, we selected a subset of \textbf{5,341} non-QSO LAEs at redshifts $z > 2.3$ with robust visually-confirmed redshifts (i.e., VI\_QUALITY $\ge2.5$).  This subset, which contains objects selected with all three ODIN narrow bands, is used for constructing the data-driven priors.

\subsubsection{Subaru Suprime-Cam LAEs}
The Suprime-Cam was a prime focus imaging camera at the 8.2-m Subaru Telescope located on Mauna Kea in Hawaii which was in operation from 1997 to 2017 \citep{Miyazaki+2002}. During this period, it conducted many narrow-band ($\approx 100$\AA) and intermediate-band (i.e., $\approx 250$\AA) filter  surveys of the extragalactic sky, resulting in discoveries of thousands of high-redshift LAE candidates \citep{ouchi_systematic_2018, Kikuta+23}.  We used photometry from five intermediate-band filters (IA427, IA464, IA484, IA505, and IA527; see Figure~\ref{fig:filters}) of $\sim$250 \AA\ FWHM from the COSMOS survey with an effective area of 2 deg$^2$ and typical depths of 25.5-26.5 mag \citep{Taniguchi+2007, Taniguchi+2015}. 

As in the case with ODIN, photometric catalogs of these data were created in combination with broad-band imaging data from the Hyper-SuprimeCam SSP (which cover the 2 deg$^2$ footprint) and selected LAE candidates as described in \citet{raichoor_DESI_HSC}. DESI targeted 1860 of these candidates during 2023, and the pipeline-processed spectra of all targeted sources were visually inspected in the same manner as described for the ODIN sources. Of these, a subset of \textbf{881} LAEs \textbf{without visually identified QSO contamination} with robust visually inspected redshifts between 2 and 4 were selected to be the test sample for this paper. 

\section{Method Overview} \label{sec:methods}
The problem of component separation, or decomposing a given dataset into constituent components, has been widely studied and is of interest to many fields. The most common method for component separation is Principal Component Analysis \citep[PCA;][]{pca}, which decomposes a dataset into orthogonal components corresponding to the directions of maximum variance. Another similar method is Independent Component Analysis \citep[ICA;][]{ica}, which returns independent (and not necessarily orthogonal) components. While often effective, these methods are ``blind" in that they assume no prior information about the data, and thus do not always identify physically meaningful components when applied to datasets in the physical sciences. Additionally, these methods do not allow for marginalization over other components when a single component is extracted.

The technique used in this work, Marginalized Analytic Dataspace Gaussian Inference for Component Separation (MADGICS), is  a covariance-based Bayesian  component separation technique \citep{Saydjari+23, Saydjari+24, MADGICS}. MADGICS has previously been used to pinpoint the wavelength and width of diffuse interstellar bands in \textit{Gaia} RVS spectra \citep{Saydjari+23} and also to improve the precision of stellar radial velocity measurements in APOGEE \citep{Saydjari+24}. 

Here we describe how we use MADGICS to decompose an LAE spectrum into its constituent components. We represent a spectrum as a length-$n$ data vector $x_{\rm tot}$, which we seek to separate into $k$ components, where $k$ is known. Each component $i = 1\dots k$ has a prior $n \times n$ covariance matrix $C_i$ which sum to $C_{\rm tot}$:
\begin{equation}\label{eq:ctot_general}
C_{\rm tot} = \sum_{i = 1}^k C_i
\end{equation}

We use the MADGICS \citep[Marginalized Analytic Dataspace Gaussian Inference for Component Separation;][]{Saydjari+23} formalism to decompose $x_{\rm tot}$ into $i = 1\dots k$  components:
\begin{equation}\label{eq:madgics}
x_i = C_i C_{\rm tot}^{-1} {x_{\rm tot}}
\end{equation}

This formalism enforces that the components $x_i$ for $i=1,\dots,k$ sum exactly to the data vector $x_{\rm tot}$. We also assume mean-zero data; in practice, the mean can be subtracted and the inference done in this space. 

A data-driven prior covariance matrix can be created for a given component as follows:
\begin{equation}\label{eq:covmatrix}
C = \frac{X X^T}{n}
\end{equation}
for an $n \times p$ data matrix $X$ consisting of $p$ length-$n$ samples of the desired component. These data vectors should be quite clean and preferably high signal-to-noise ratio (SNR), or the covariance estimate will be contaminated and the statistical power of the approach reduced.\footnote{Equation~\ref{eq:covmatrix} is an unbiased but noisy estimator of C, which is a biased estimator of $C^{-1}$, see \citet{Hartlap_2006} for more information.}

We can evaluate how well a spectrum $x_{\rm tot}$ is expressed by a set of components by evaluating its $\chi^2$ metric, where:
\begin{equation}
\chi^2 = x_{\rm tot}^TC_{\rm tot}^{-1}x_{\rm tot}
\label{eq:chisq}
\end{equation}

This generalized method can be applied to LAE spectra as follows: here we model an LAE spectrum as a sum of three components: an LAE spectrum, a sky residual spectrum, and a residual spectrum. Thus $k = 3$, every $x_{\rm tot}$ is an LAE spectrum, $x_i$ is the extracted component, where $i$ corresponds to sky residual, LAE, or residual, and:
\begin{equation}
C_{\rm tot} = C_{\rm sky} + C_{\rm LAE} + C_{\rm res}
\label{eq:Ctot}
\end{equation}

To evaluate the effectiveness of adding an LAE component at redshift $z$, we calculate the difference in $\chi^2$ resulting from including that component:
\begin{equation}
\Delta \chi^2(z) = \chi^2 (\text{sky + LAE($z$) + res}) - \chi^2 (\text{sky + res})
\label{eq:deltachisq}
\end{equation}
This $\Delta \chi^2$ should be negative, as including an LAE component can only increase the goodness of fit, yielding a lower $\chi^2$ value.

\subsection{Prior generation}\label{sec:priors}
\subsubsection{Sky residual prior}
We start with sky residual spectra, which have already been sky-subtracted by the DESI pipeline, but have sky residual lines that require additional marginalization. To create our sky residual prior, we start with 19,989 sky residual spectra from exposures used to observe the ODIN LAE targets. The sky residual spectra do not contain the \lya or any other target lines. We interpolate both the flux values and inverse variances reported by the DESI pipeline for these spectra to be log-spaced with a spacing of $\Delta \ell = \Delta \log_{10} \lambda = 5 \times 10^{-5}$ (where $\lambda$ is in \AA), as log-$\lambda$ binned spectra allow us to represent approximate redshift changes as pixel shifts. We then remove outlier spectra that sum to greater than $10^6$ or have more than 1,000 masked pixels, leaving 19,657 spectra. 

To ensure that our prior captures most of the variance from the sky residual component, we manually mask the most prominent sky residual lines. Any wavelength bin that has more than 1/3 of the sky residual spectra above the 95th or below the 5th percentile is masked, along with 3 pixels on each side. In practice, 45 wavelength bins out of 8,720 meet this criteria and 89 total are masked. The sky residual prior is then created without using these wavelength bins, and in any reconstructions all flux at these wavelengths is assigned to the sky residual component.

We express our sky residual prior as a full-rank data-driven covariance matrix using Equation~\ref{eq:covmatrix}. Due to increased noise and reddening at shorter (bluer) wavelengths, we fit a fourth-order polynomial $y$ to the log of the diagonal of this covariance matrix, with units of $\log({\rm flux^2})$: 
    
\begin{equation}\label{eq:sky_poly}
    \begin{split}
    y(\ell) = &686.2\ell^4 - 10383.1\ell^3 +  58903.2\ell^2 \\ 
   &- 148486.2\ell + 140343.7
    \end{split}
\end{equation}
where $\ell$ is log-wavelength in $\log_{10}$(\AA).
This yields a flux rescaling of:
\begin{equation}\label{eq:flux_correction}
    f'(\ell) = \frac{f(\ell)}{\sqrt{10^{y(\ell)}}}
\end{equation}
which we apply to all sky residual spectra. This also necessitates a corresponding inverse variance $i$ rescaling of:
\begin{equation}\label{eq:ivar_correction}
    i'(\ell) = 10^{y(\ell)}i(\ell)
\end{equation}
Normalizing the spectra in this way ensures that the eigenvectors of our covariance matrix will be dominated by the sky residual lines of interest instead of noise at low redshifts. We also apply this flux and inverse variance correction for all target spectra, and perform our joint inference of the components and redshift determination in this transformed space.

 We then use these rescaled spectra to create a new data-driven covariance matrix. From there, the top 50 eigenvectors\footnote{We also did the analysis with 25 and 100 eigenvectors, and found that performance did not improve beyond 50. We are incentivized to minimize the number of eigenvectors to optimize computational speed.} are used to create a low-rank approximation of the covariance matrix. Thus, we express the sky residual covariance $C_{\rm sky} = VV^T$, where $V$ is the matrix of the top 50 eigenvectors scaled by the square root of their associated eigenvalues. This $C_{\rm sky}$ serves as our prior for the sky residual component. 

\begin{figure}
    \centering
    \includegraphics[width=\linewidth]{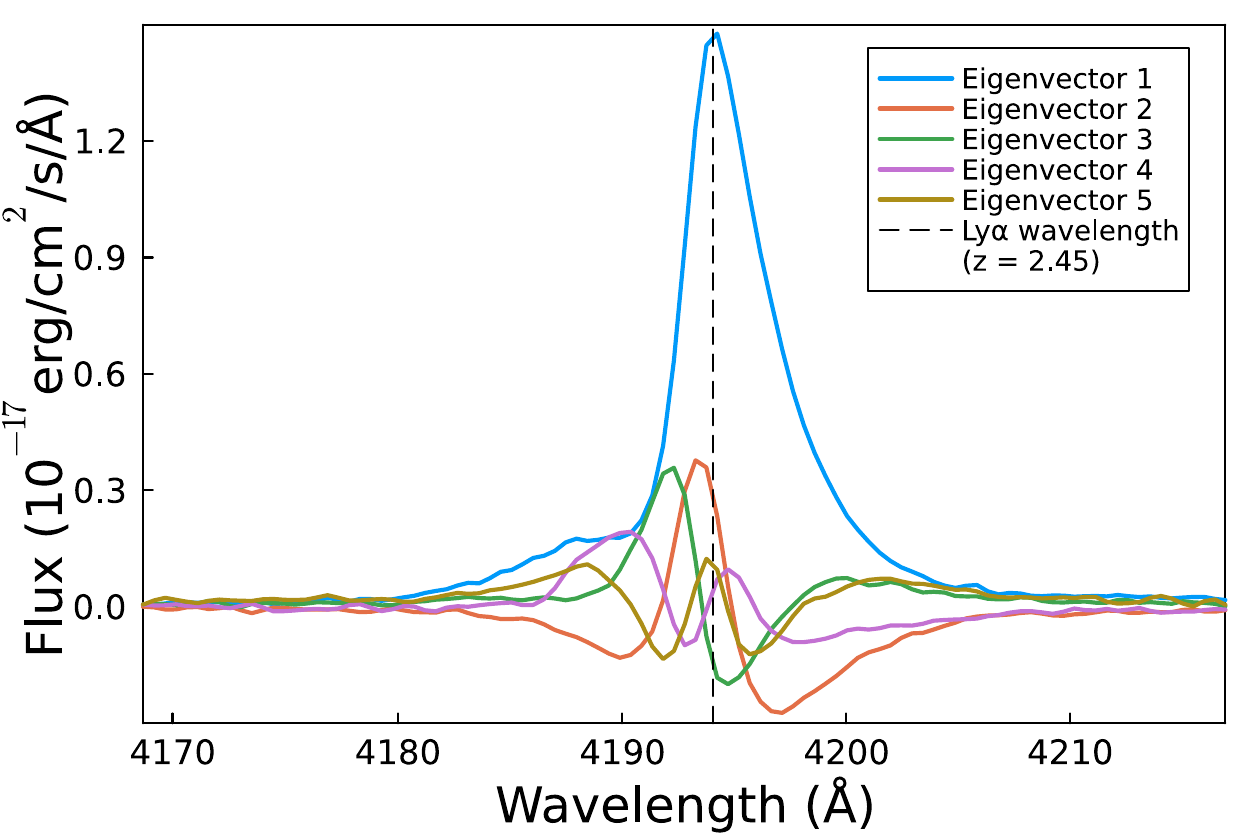}
    \caption{Top five eigenvectors from the data-driven covariance matrix of ODIN LAE targets shifted to be at $z = 2.45$. Eigenvectors are scaled by the square root of their respective eigenvalues. }
    \label{fig:lae_eigenvecs}
\end{figure}

\subsubsection{LAE prior}

To create the LAE prior, we use DESI spectra from ODIN LAE candidates with VI redshifts between 2.3 -- 4.5. Here we consider the VI redshifts to be ground truth, although in practice there is likely a small error rate for both LAE identification and redshift precision. In this work, we only use LAE target spectra with an average ``VI\_QUALITY" score $>$ 2.5 (on a scale from 0 to 4), and which \textbf{were not visually identified as QSOs,} as discussed in Section~\ref{sec:odin_laes}.

These spectra interpolated to the same log-spacing as the sky residual spectra ($\Delta \ell = 5 \times 10^{-5}$). They are then scaled as in Equation~\ref{eq:flux_correction} and decomposed into sky residual and residual components using Equation~\ref{eq:madgics}. We use the residual component from this decomposition so as to not include sky residual lines in the prior covariance matrix. To ensure that we use high SNR spectra, we calculate the variance of the flux values of the spectrum and include only those with variances below the 75th percentile of the whole population to exclude overly noisy spectra, leaving \textbf{4,005} spectra. Given the visually inspected redshifts, we shift the remaining spectra to all be at $z = 2.45$, the central redshift of the N419 ODIN filter. 

We can express our data as an $n \times p$ data matrix containing $p$ spectra each with $n$ wavelength bins, where masked values are set to zero. If we define the $n \times p$ matrix of residual ODIN LAE spectra (after decomposition of the sky residual component) rescaled as in Equation~\ref{eq:flux_correction} as $D$, we can define a matrix $A$ such that:
\begin{equation}
    A_{ij} = \begin{cases}
        1 & D_{ij} \neq 0 \\
        0 & {\rm otherwise}
    \end{cases}
\end{equation}
where each element corresponds to the number of data vectors with non-masked flux values in wavelength bin $j$. We can then define the data-driven LAE prior covariance as:
\begin{equation}
    C_{\rm LAE} = \frac{DD^T}{AA^T}
\end{equation}
where here division denotes element-wise matrix division. This is a generalization of Equation~\ref{eq:covmatrix} where each element of the covariance matrix is being divided by the number of data vectors with non-masked values in that wavelength bin, instead of dividing the entire matrix by the total number of data vectors.

Figure~\ref{fig:lae_eigenvecs} shows the top five eigenvectors from the LAE prior covariance matrix $C_{\rm LAE}$, scaled by the square root of their respective eigenvalues. The \lya line at $z = 2.45$ is at 4194 \AA, and so $C_{\rm LAE}$ is defined between 4133 and 4278 \AA\ to more than encompass the entire shape of the \lya line. Physically interpretable features of the \lya line, such as its blue cutoff and red wing, are reproduced in these eigenvectors by our totally data-driven method. These eigenvectors can be easily shifted to any redshift within the given wavelength range to perform component separation across a grid of redshifts. The ability of our method to fit multiple eigenvectors allows it to detect various possible shapes of the \lya line. In practice, we express the LAE prior covariance matrix as a low-rank approximation using the top two eigenvectors to maximize computational speed. 


  For these spectra, redshifts are determined based on the observed \lya emission line, which can be offset from the systemic redshift of the galaxy by 100 -- 300 km/s \citep{verhamme_recovering_2018, Steidel+18, Pahl+21a}; this should not significantly impact cosmological analyses done at large length scales, but may impact small-scale analyses.

\subsection{Spectroscopic redshift determination}

For each target spectrum, we scan over redshifts ranging from $z = 2$ to $z = 4$ with 4,501 steps of 1 log-$\lambda$ pixel each, with the average pixel corresponding to a redshift step of $\Delta z = 4.5 \times 10^{-4}$. The sky residual and LAE prior covariances are the same across all targets, with the LAE prior changing for each redshift value. For each target, we create a unique residual prior covariance matrix $C_{\rm res}$ using the per-target variances at each wavelength (inverting the inverse variances at each wavelength reported by the DESI pipeline, and scaled as in Equation~\ref{eq:ivar_correction}) on the diagonal of an otherwise empty matrix. 

\begin{figure}
    \centering
    \includegraphics[width=\linewidth]{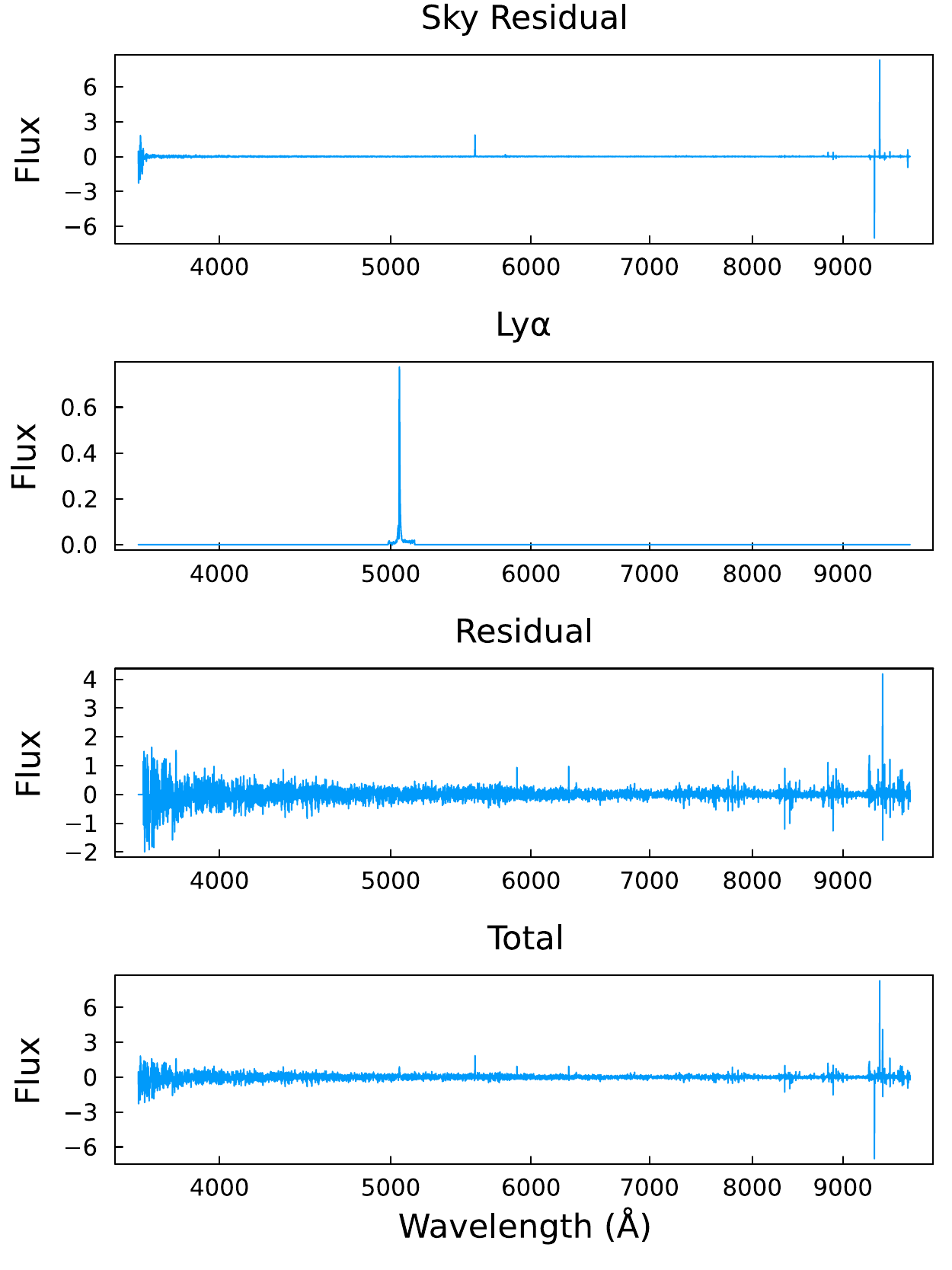}
    \caption{An example decomposition of a $z \approx 3.16$ LAE spectrum into its constituent sky residual, LAE, and residual components. Note the different vertical scaling in each panel.}
    \label{fig:separation}
\end{figure}

At each redshift step, we calculate the \dchisq{} as in Equation~\ref{eq:deltachisq} and find the redshift that minimizes this \dchisq{} profile, which indicates that including an LAE component at that redshift improves the fit the most. The $\Delta \chi^2$ also indicates a level of confidence or SNR in our detection; higher $|\Delta \chi^2|$ indicates a more robust detection of an LAE. To achieve a more precise redshift estimate, we do an additional, finer scan along $\pm$ 5 pixels around this best redshift value in steps of 0.1 pixels, and assign the target the minimum redshift from this final scan. Averaged over the \textbf{881} Subaru targets, it took $\sim0.4$ seconds to determine the redshift for one target spectrum.

For information on computational speedups used to evaluate the $\Delta \chi^2$ quickly, see Appendix A.

\subsection{Uncertainty quantification}\label{sec:uncertainty}
For each target spectrum, by scanning over a grid of redshifts, we create a \dchisq{} profile as a function of redshift, and assign each target the redshift value that minimizes this profile. Redshift uncertainties can be derived for each target by assessing the curvature of the \dchisq{} profile around the minimum. A steeper curve indicates less uncertainty than a wider curve. The \dchisq{} profile corresponds to the negative log-likelihood function multiplied by a factor of 2, and to approximate the second derivative, we fit a parabola to this log-likelihood surface $\pm$ 5 pixels (of the finer 0.1 pixel sampling) around the minimum pixel $p$ and measure its width. This yields an uncertainty in pixel space $p_{unc}$, which we then propagate to a redshift error $z_{unc}$ using
\begin{equation}
    z_{unc} = \frac{d z}{d p} p_{unc} = z\frac{d \ln z}{d p} p_{unc} =  10^{ \Delta\ell p} \Delta \ell (\ln 10 ) p_{unc}
\end{equation}

where $\Delta \ell = 5 \times 10^{-5}$ is the log-wavelength pixel spacing.

\section{Results \& Validation} \label{sec:results}
With priors in hand, we can use the MADGICS formalism (Equations~\ref{eq:ctot_general} and ~\ref{eq:madgics}) to decompose an LAE spectrum into sky residual, LAE, and residual components. Figure~\ref{fig:separation} shows an example decomposition of an LAE spectrum. MADGICS guarantees that the components sum exactly to the data.

\subsection{Redshift determination for Subaru LAE targets}
We use the priors described in Section~\ref{sec:methods} to decompose and determine redshifts for DESI spectra of \textbf{881}  LAE candidates targeted with Subaru that have VI redshifts between 2 and 4. We evaluate our method on a held-out test set, meaning that none of the spectra for which we show results were used to create the LAE prior covariance matrix.

\begin{figure}
    \centering
    \includegraphics[width=\linewidth]{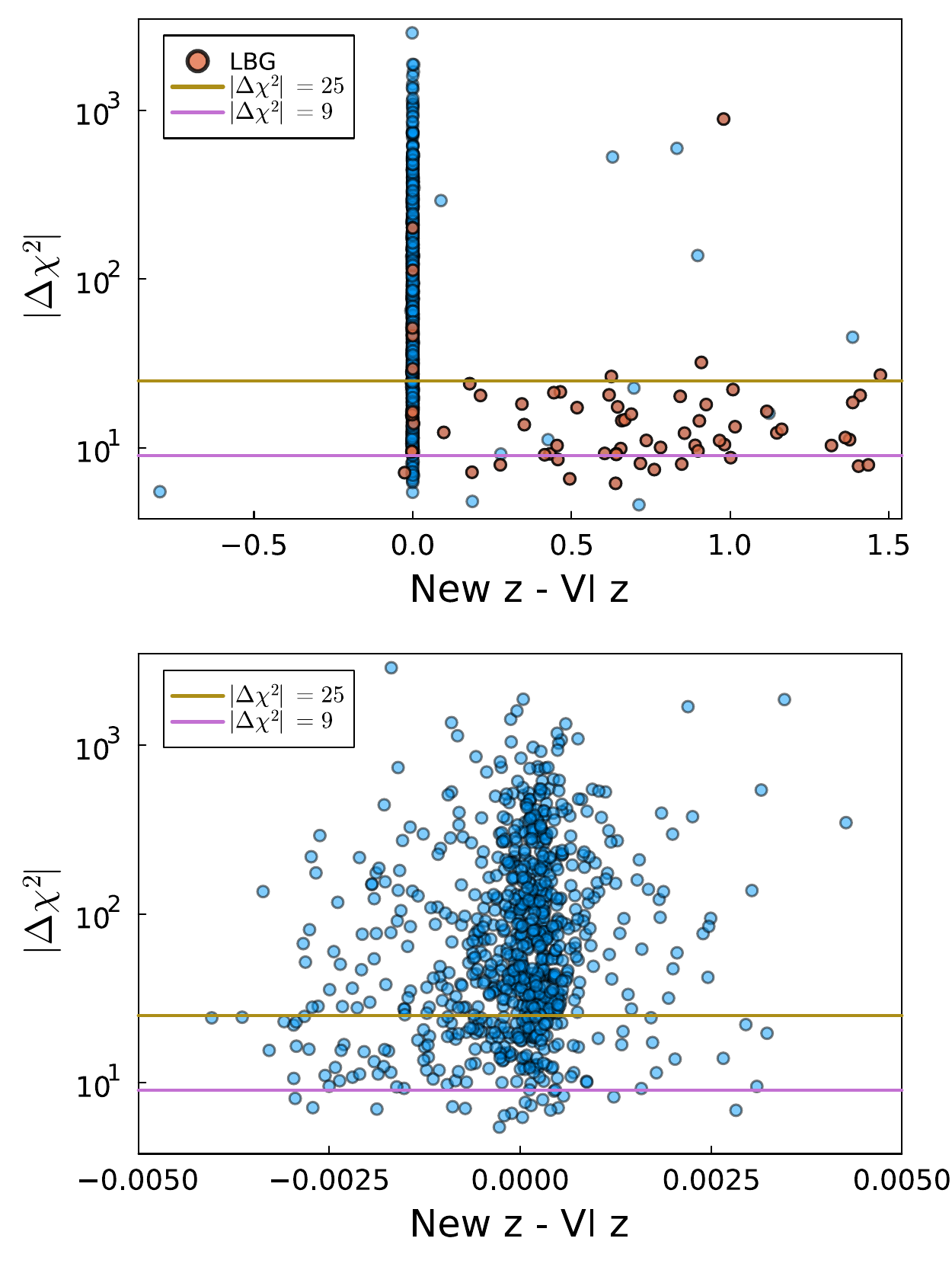}
    \caption{$|\Delta \chi^2|$ vs. redshift residual (new $z$ - VI $z$) for \textbf{881} Subaru LAE targets that were not used to create the LAE prior (blue points). (top) Orange points denote LAE targets with ``LBG" in the VI comments. (bottom) The same plot shown only between -0.005 $<$ New z - VI z $<$ 0.005. \textbf{812 (92.3\%)} of the \textbf{881} Subaru LAE targets are contained within this range.}
    \label{fig:chisq}
\end{figure}

Figure~\ref{fig:chisq} shows the $|\Delta \chi^2|$ (calibrated as described in Section~\ref{sec:error_calib}) as a function of redshift residual (new - VI $z$) for this test set. There is generally good agreement between the redshifts determined by our pipeline and those from visual inspection: \textbf{812 (92.3\%)} of the \textbf{881} evaluated spectra were determined by our method to have a redshift within 0.005 of the VI. Most disagreements between the two methods occur at low $|\Delta \chi^2|$, which would indicate low SNR or an uncertain detection. Several factors could contribute to our pipeline determining redshifts outside of this threshold: some of the sources are Lyman Break Galaxies (LBGs) or other objects which do not have a strong Ly$\alpha$ emission line; other objects that show strongly double-peaked \lyab, or very weak lines; and some could be erroneous redshifts reported by the visual inspection.  

\begin{figure}
    \centering
    \includegraphics[width=\linewidth]{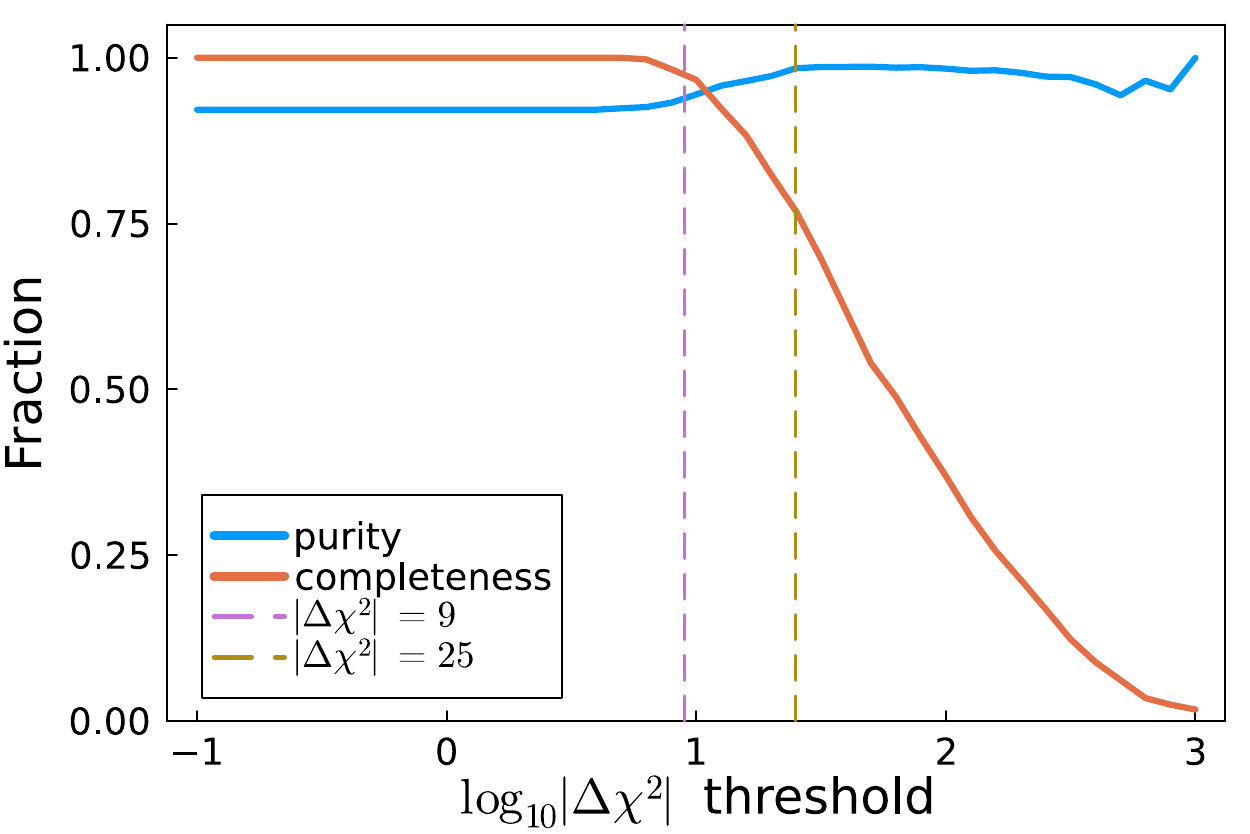}
    \caption{Purity \& completeness as a function of $|\Delta \chi^2|$ cutoff for determining if a target is an LAE. Here a redshift is ``correct" if it is within 0.005 of the VI redshift. Note that we are evaluating our method on a pre-selected sample of solely LAE targets. Pink dashed line indicates $|\Delta \chi^2| = 9$, and gold dashed line indicates $|\Delta \chi^2| = 25$.}
    \label{fig:purity}
\end{figure}

To create a binary classifier (LAE or not LAE), one could impose a threshold \dchisq{} value, above which any target is classified as being an LAE. Figure~\ref{fig:purity} shows the purity and completeness of the LAE sample as a function of the potential calibrated \dchisq{} threshold value, where a ``correct" detection is a target that has a pipeline redshift within 0.005 of its VI redshift. Note that we are evaluating our method on a pre-selected sample of solely LAE targets with no known non-LAEs, so a high purity across all thresholds is to be expected. Figures~\ref{fig:chisq} and~\ref{fig:purity} suggest that a value of $|\Delta \chi^2|~\approx$~25 could be an effective threshold value for classifying an object as an LAE, which roughly corresponds to a 5$\sigma$ cut.

In addition to the \dchisq{} value, which denotes the improvement in the fit from adding an LAE component, the $\chi^2 (\text{sky + LAE($z$) + res})$, hereafter referred to as the ``total $\chi^2$", indicates how good of a fit the entire model is to the data. Targets that are not LAEs but have other emission lines, such as additional sky residual lines or low-redshift interlopers, may have a large \dchisq{} value, but will also likely have a large total $\chi^2$ value since our LAE component does not represent their emission line well. This total $\chi^2$ value can be used to identify potential false positives.

\begin{figure}
    \centering
    \includegraphics[width=\linewidth]{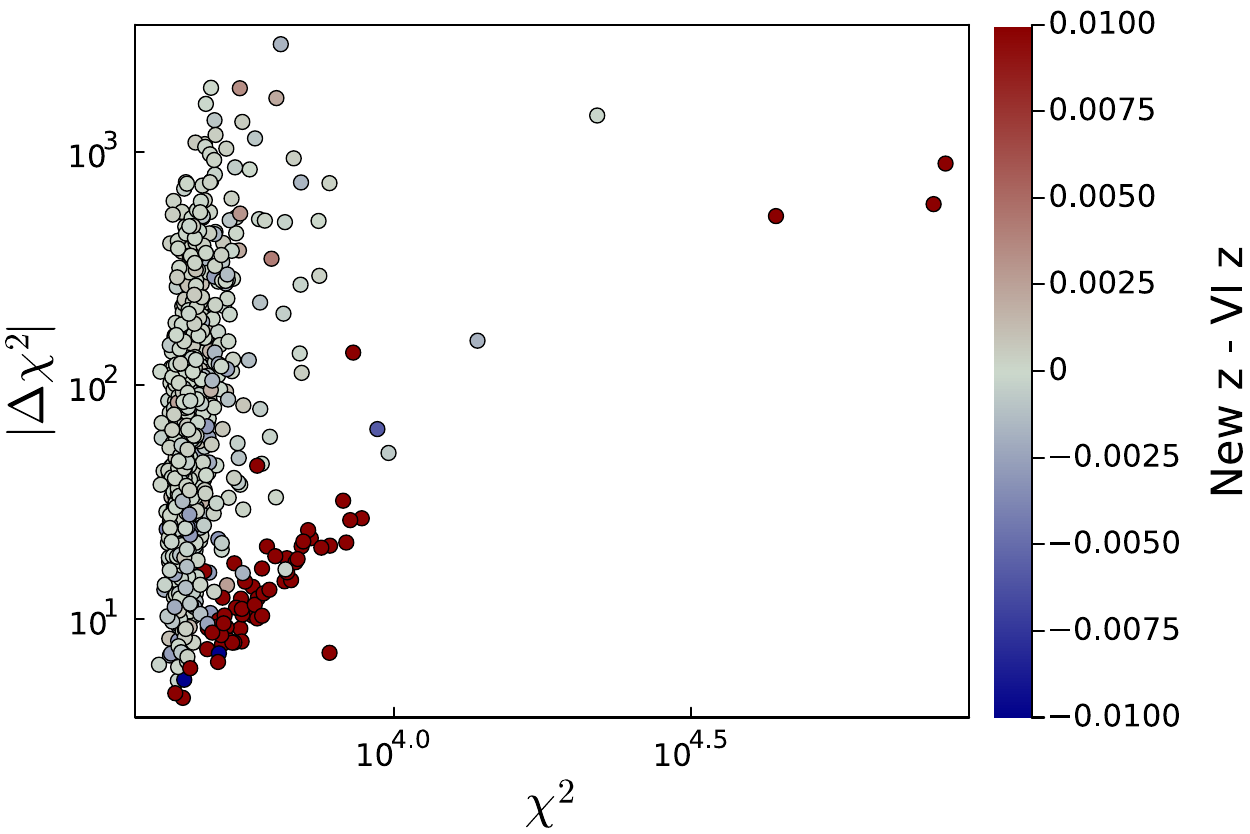}
    \caption{$\Delta \chi^2$ vs. $\chi^2 (\text{sky + LAE($z$) + res})$, colored by redshift residual (new z - VI z), for \textbf{881} Subaru LAE targets. Colorbar limits are set to $\pm$0.01, with redshift residuals beyond those values set to the darkest colors.}
    \label{fig:delta_vs_chisq}
\end{figure}

Figure~\ref{fig:delta_vs_chisq} shows the \dchisq{} and total $\chi^2$ values for the \textbf{881} Subaru LAE targets, colored by redshift residuals. For this dataset, there are \textbf{five} outliers with total $\chi^2 > 10^4$. Visual inspection of these objects shows that they include quasar contamination and/or potential cosmic rays. Of these \textbf{five} outliers, our pipeline still identifies a redshift within 0.005 of the VI redshift for \textbf{two} of them.

As seen in Figure~\ref{fig:chisq}, the majority of poor redshift fits can be identified by imposing a \dchisq{} cut. However, false positives can be further identified by imposing a total $\chi^2$ cut, drawing a vertical line on Figure~\ref{fig:delta_vs_chisq}. Additionally, a diagonal line can be drawn in this space to cleanly separate out the majority of redshift failures. A cut based on the total $\chi^2$ is likely useful for future applications of this method to larger, less pure datasets.

\subsection{Injection-recovery tests}\label{sec:simulations}
\subsubsection{Uniform noise}\label{sec:uniform_noise}

To better quantify the performance of our method, we can perform injection-recovery tests with simulated LAE spectra where we inject a \lya line at a known redshift and recover the redshift with our pipeline. Unlike real LAE spectra, for which we depend on visually inspected redshifts, these simulated spectra have perfectly known redshift values with no uncertainty.

We create a simulated LAE spectrum by injecting a normalized median spectrum from targets selected with the ODIN N501 filter (scaled as in Equation~\ref{eq:flux_correction}) at a randomly generated redshift and adding noise. To mimic the shape and SNR of real LAE spectra, the simulated spectra are made using a randomly sampled true redshift $z$ and scaling factor $\eta$:
\begin{equation}\label{eq:redshift}
z \sim \mathcal{U}(2,4)
\end{equation}
\begin{equation}
\eta \sim \mathcal{U}(0, 50)
\end{equation}

The constant scaling factor $\eta$ scales the injected LAE signal. A simulated spectrum $\mathbf{m}$ is then drawn from a multivariate normal distribution with variance $\sigma^2 = 0.3^2$ for each wavelength:
\begin{equation}
\mathbf{m}\sim \mathcal{N}(\eta \: \text{LAE}(z), 0.3^2 \mathbb{I})
\end{equation}
where $\text{LAE}(z)$ denotes the injected LAE signal, shifted to redshift $z$.

\begin{figure}
    \centering
    \includegraphics[width=\linewidth]{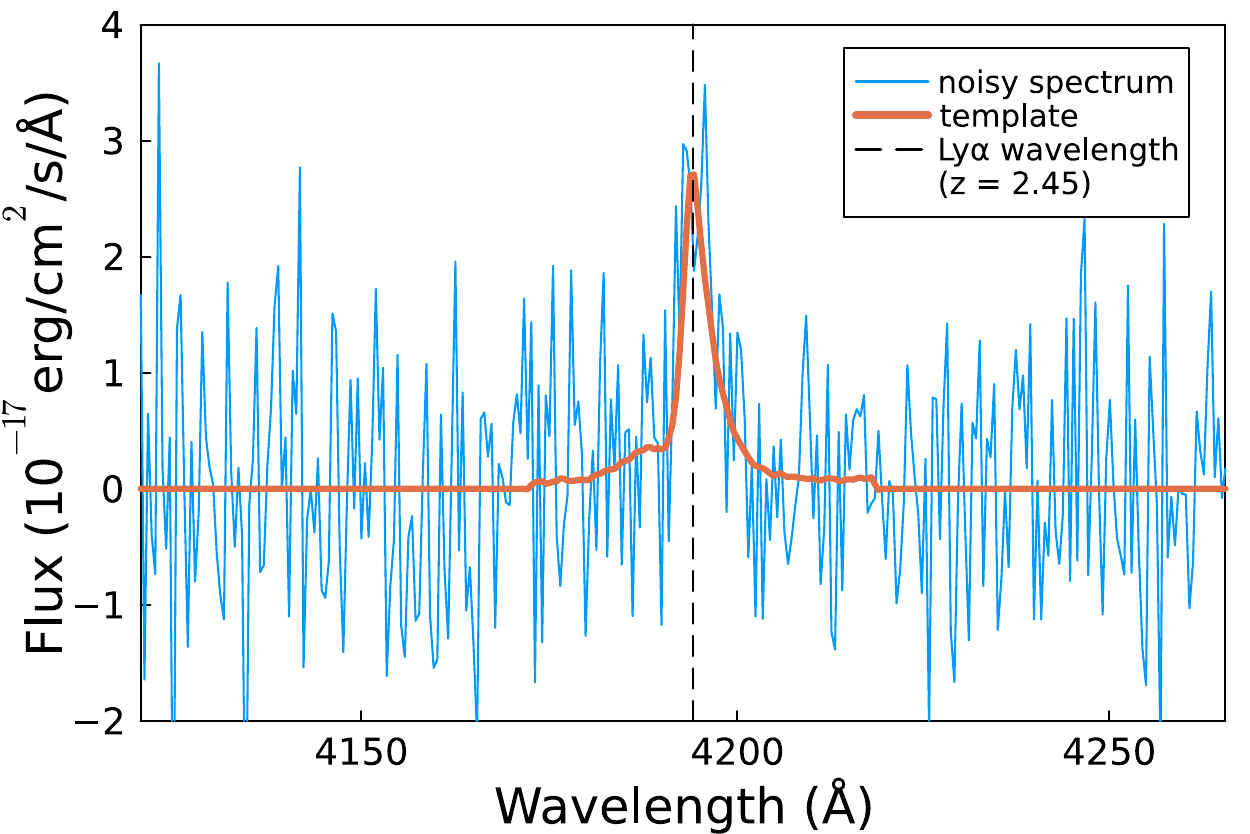}
    \caption{An example simulated spectrum of an LAE at $z = 2.45$, zoomed in around the \lya line. A stacked ODIN LAE spectrum (orange) is injected at a known redshift (black dotted line), and Gaussian noise is added to create a noisy spectrum (blue). This example has SNR $\approx$ \textbf{8}, calculated using Equation~\ref{eq:snr}.}
    \label{fig:noisy}
\end{figure}

Using this procedure, we create \textbf{5,000} simulated LAE spectra and use our pipeline to automatically determine their redshifts using $0.3^2 \mathbb{I}$ as $C_{\rm res}$ and no sky component. An example simulated LAE spectrum can be seen in Figure~\ref{fig:noisy}. For each simulated spectrum, we inject a \lya{} emission line at a randomly selected redshift and calculate the signal-to-noise ratio (SNR) as:

\begin{equation}\label{eq:snr}
{\rm SNR} = \frac{\eta}{\sigma \sqrt{A}}
\end{equation}
where
\begin{equation}
    A = \frac{1}{\int p^2 d\lambda}
\end{equation}
where $p$ is the injected signal and $\sigma$ is the standard deviation of the constant noise added to the injected signal (here 0.3), following the methodology of \citet{Portillo+20}.

\subsubsection{Injection into sky fibers}
\label{sec:real_injection}

To better approximate real LAE spectra observed with DESI, we can again inject a normalized LAE signal at a known redshift into actual DESI sky residual spectra and examine how well our method can recover them.

For this, we randomly sample a redshift as in Equation~\ref{eq:redshift} and inject the LAE signal scaled by a constant $\eta$ into a random sky residual spectrum from the selection outlined in Section~\ref{sec:priors}, ensuring proper scaling using Equations~\ref{eq:flux_correction} and \ref{eq:ivar_correction} for inference in the transformed space. For this case, we calculate $\sigma$ as the robust standard deviation (IQR/1.34896) of the sky residual spectrum between 3647 and 6078 \AA, corresponding to the \lya wavelength at $2 < z < 4$. Here, the scaled variance from the sky spectrum is used on the diagonal of the residual covariance ${\rm C_{res}}$. 

\begin{figure}
    \centering
    \includegraphics[width=\linewidth]{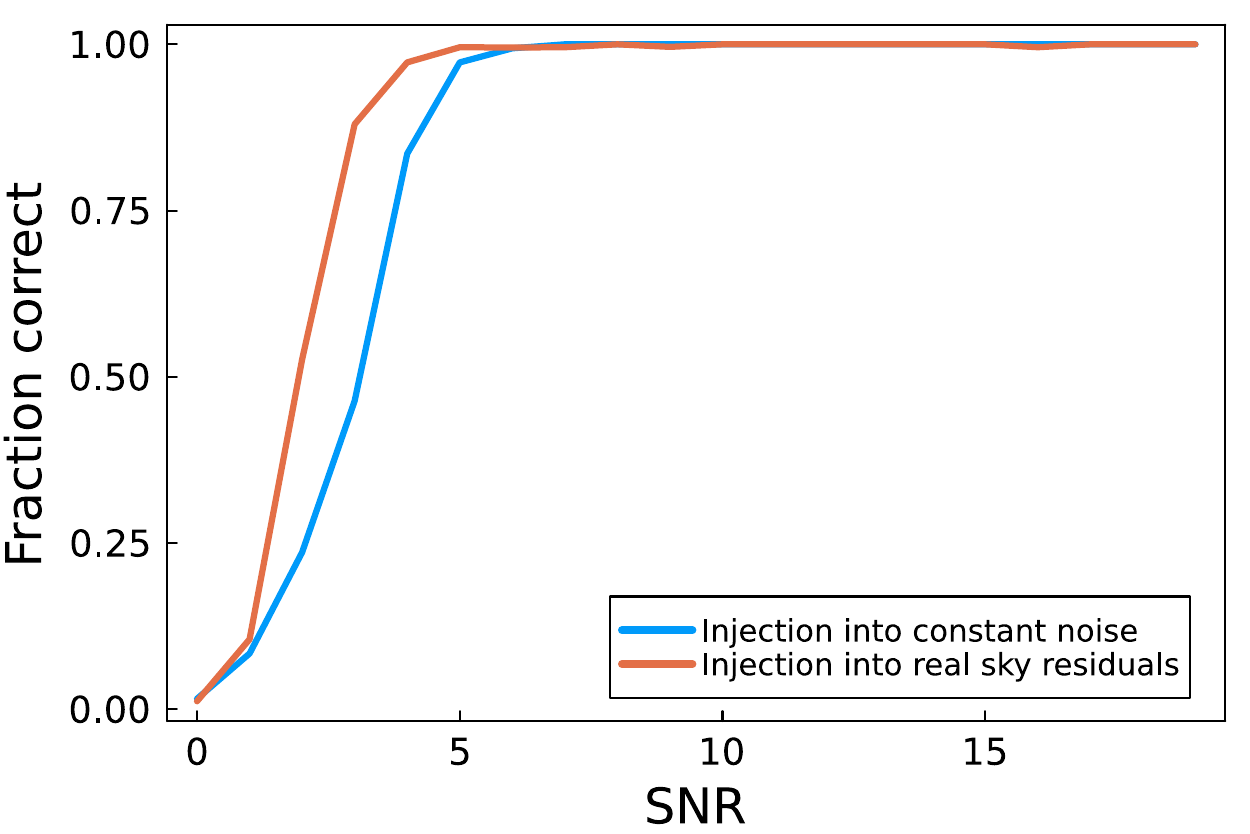}
    \caption{Fraction of correctly recovered redshifts (within 0.005 of injected redshift) as a function of SNR for \textbf{5,000} simulated spectra created by injecting a median LAE spectrum into uniform noise (blue line, as described in Section~\ref{sec:uniform_noise}) and into real sky residual spectra (orange line, as described in Section~\ref{sec:real_injection}). }
    \label{fig:snr}
\end{figure}

For both kinds of injection tests (constant noise and sky residual spectra), we then bin the simulated spectra in SNR bins of width 1 ranging from SNR = 0 -- 25. Within each bin, we calculate the fraction of simulated spectra for which our method correctly recovered the injected redshift, determining a redshift estimate to be ``correct" if it is within 0.005 of the true injected redshift regardless of its \dchisq{} value. Figure~\ref{fig:snr} shows the fraction of correctly recovered redshifts as a function of SNR. For the constant-noise tests, 50\% of simulated LAE redshifts were correctly identified at SNR $\approx$ 3, and 100\% at SNR  $\approx$ 6, while for the injection into real sky residuals, these values occur at SNR $\approx$ 2 and 5, respectively. 

\subsubsection{Error calibration}
\label{sec:error_calib}
The \dchisq{} and redshift uncertainty measurements are made under the assumption that the noise and uncertainties in the data are well modeled and a mean zero prior on each component. The model mismatch caused by assuming a mean zero prior for an emission line can distort the $\Delta\chi^2$ surface we measure as a function of redshift. In addition, there are many factors, such as stochasticity in the data as well as over- or under-estimation of the intrinsic noise, that could result in miscalibrated redshift errors from our pipeline. To correct for this  $\chi^2$ surface distortion and calibrate the redshift errors we use our simulated LAE spectra created for injection tests to examine the calibration of our uncertainties and calculate a correction factor to be applied to the real data. For both kinds of injection tests, we generate \textbf{5,000} new simulated spectra with a higher SNR range (SNR $\sim 100$), and fit them with our pipeline as described above. For each simulated spectrum, the z-score was calculated as follows:

\begin{equation}
    z_{\rm score} = \frac{{\rm fit}\; z - {\rm true} \;z}{z_{unc}}
\end{equation}
where $z_{unc}$ is here the uncertainty on the redshift estimate as outlined in Section~\ref{sec:uncertainty}. Well-calibrated error bars yield z-scores that follow a unit normal distribution. To be robust to outliers, we evaluate this by calculating the interquartile range (IQR) of the distribution of z-scores and comparing it to the IQR of standard normal distribution (1.34896).

We calibrate our $z_{unc}$ values by assuming that the $\Delta\chi^2$ surface distortion is a simple multiplicative correction, such that if the amplitude of the $\Delta\chi^2$ value is corrected, then the curvature will also be corrected. This allows us to derive a correction factor and independently verify that the resulting $z_{\rm score}$ for the updated $z_{unc}$ values is in fact well-described by a unit normal distribution. To correct the amplitude of the $\Delta\chi^2$ surface, we assume that the $\Delta\chi^2$ is dominated by removing the LAE signal which would have otherwise factored into the residuals. Then, the $\sqrt{(\Delta\chi^2)}$ should be equal to the SNR, as defined in Section~\ref{sec:uniform_noise}. We can compute the multiplicative factor by which this differ on injection tests, which we call $s$, as:

\begin{equation}
    s = \frac{\sqrt{|\Delta \chi^2|}}{{\rm SNR}}
\end{equation}
In practice, we take $s$ to be the median ratio across all simulated spectra. Once $s$ is calculated, the $z_{unc}$ and  $\Delta \chi^2$ are recalibrated as follows:
\begin{equation}
    z_{unc}' = sz_{unc}
\end{equation}
\begin{equation}
    |\Delta \chi^2|' = \frac{|\Delta \chi^2|}{s^2}
\end{equation}

\begin{figure}
    \centering
\includegraphics[width=\linewidth]{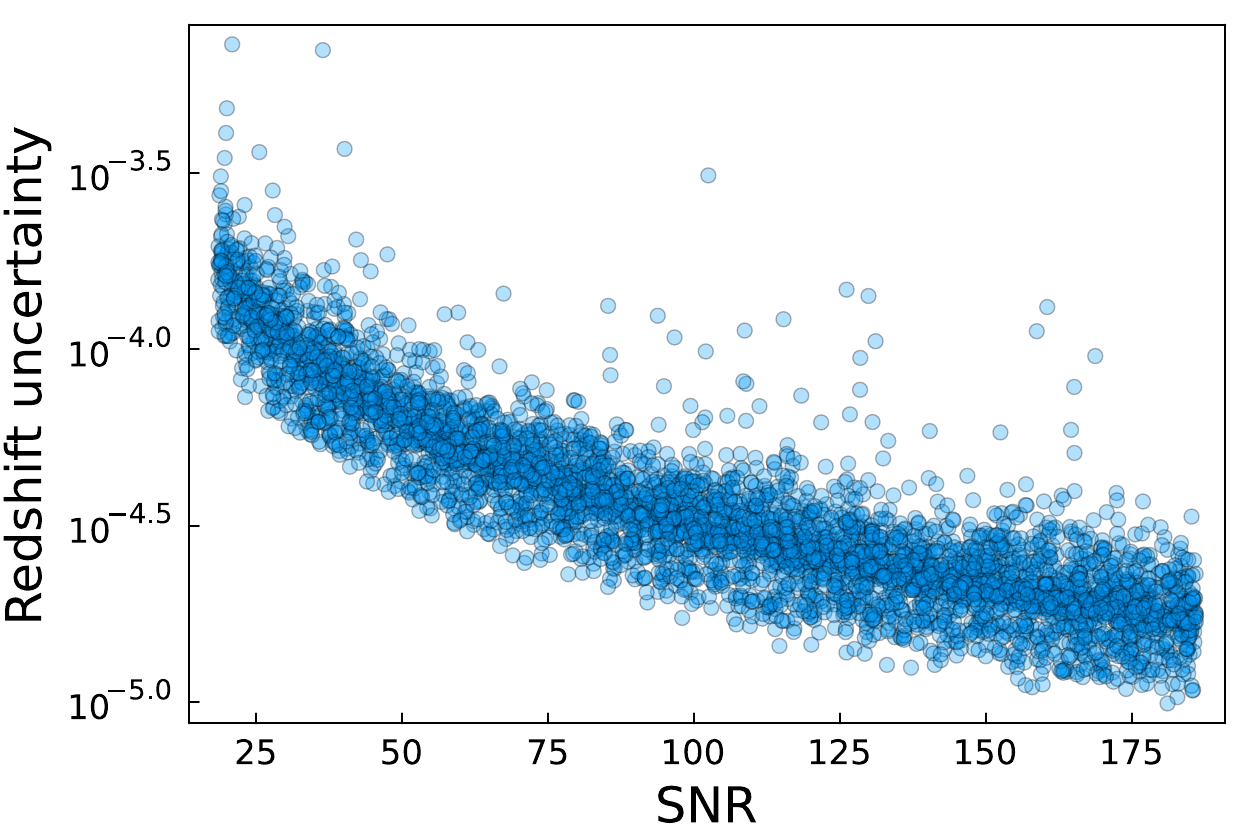}
    \caption{Redshift uncertainty (calculated as described in Section~\ref{sec:uncertainty}) as a function of SNR for \textbf{5,000} simulated spectra created by injecting a median LAE spectrum into real sky residual spectra.}
    \label{fig:redshift_uncertainty}
\end{figure}

For the constant-noise injection tests, we calculate $s = 0.97$, suggesting that the noise was well-modeled, and the adjusted z-score distribution yielded an IQR of \textbf{1.3276/1.34896 = 0.984}. For the injection into real sky residual spectra, we calculate $s = 2.16$, and the adjusted IQR is \textbf{1.2426/1.34896 = 0.921}. Both ratios are quite close to 1, suggesting that after calibrating the $\Delta \chi^2$ surface, the adjusted redshift errors are also well-calibrated.

Figure~\ref{fig:redshift_uncertainty} shows the redshift uncertainty as a function of SNR for the simulated spectra created by injection into real sky residual spectra. As expected, a higher SNR detection leads to a more precise redshift estimate, and all redshift uncertainties are less than one average pixel size.  

\begin{figure*}
    \centering
    \includegraphics[width=\linewidth]{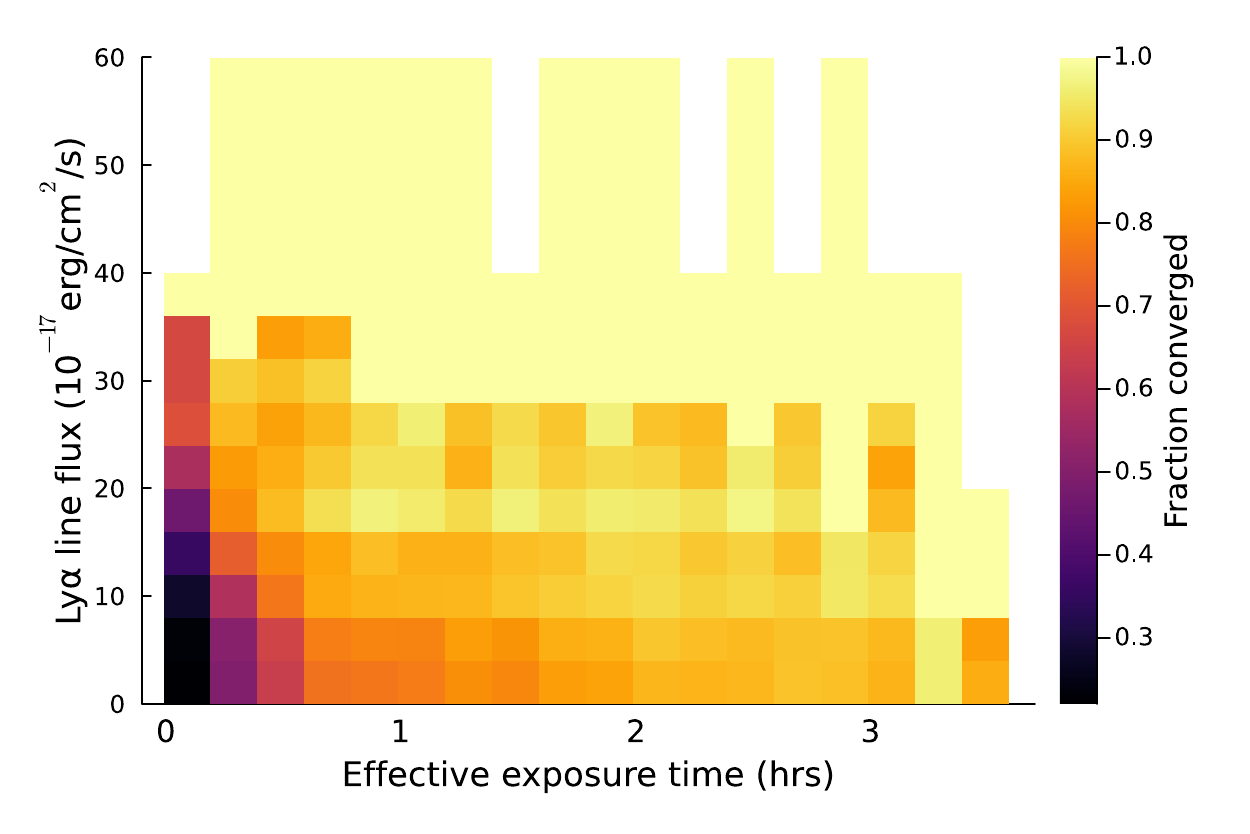}
    \caption{Cumulative fraction of coadded spectra with converged redshift estimates ($\sigma_z < $ 0.001) binned over exposure time and \lya line flux for \textbf{310} Subaru LAE targets observed with $\geq$ 15 exposures. Each bin shows the fraction of objects with ($\sigma_z < $ 0.001) within that exposure time bin and with \lya line flux greater than or equal to the value denoted by each bin.}
    \label{fig:exposures}
\end{figure*}

We can use the $s = 2.16$ factor calculated from the injection tests on real sky residual spectra to calibrate the reported \dchisq{} and redshift uncertainties on the real LAE target spectra, using the VI redshifts as ground truth. For our dataset targeted with the Subaru Suprime-Cam, we apply the same corrections and recover an adjusted IQR of \textbf{2.0214/1.34896 = 1.498}, where we exclude catastrophic outliers ($|\Delta z| > 0.005$) from the IQR calculation to reduce bias. 

This $s$ factor is used with Equations 20 and 21 to recalibrate the \dchisq{} and redshift uncertainty for an LAE detection for the Subaru LAE candidates, as seen in Figures~\ref{fig:chisq} and~\ref{fig:purity}. A larger IQR ratio is expected for real data, which includes uncertainties associated with the VI redshifts as they are not ground truth. We can also fit an additional uncertainty term associated with the VI redshifts $z_{unc, \rm VI}$ that is added in quadrature with the redshift uncertainties. Fitting this to an IQR ratio of 1 leads to an approximate VI redshift uncertainty of $z_{unc, \rm VI} \approx 3 \times 10^{-4}$. This provides an approximate quantification of the potential uncertainties of the VI redshifts. Including this VI uncertainty in $z_{unc}$, we find that our pipeline determines redshifts within 3$z_{unc}$ of the VI redshift for \textbf{720} out of the \textbf{881} Subaru LAE targets \textbf{(81.7\%)}. This provides an additional metric of redshift accuracy of our pipeline, in addition to the previously reported accuracy of $>$90\% of Subaru LAE targets being within 0.005 of the VI redshift.

\subsection{Implications for DESI-2}\label{sec:desi_ii}

In advance of DESI-2, in addition to scalable redshift analyses, it is important to understand which objects to target and for how long to observe them in order to maximize the scientific output of the survey. This work has three main implications for DESI-2:
\begin{itemize}
    \item We develop a scalable automatic redshift determination tool to enable analysis of thousands of LAE spectra.
    \item We present a framework for calculating a \dchisq{} value that can serve as a proxy for confidence that a given target is an LAE. This   can be used to inform targeting of photometric candidates selected from medium-band imaging and determine the optimal observing parameters for spectroscopy.
    \item We assess how much exposure time is necessary to converge on a redshift estimate for LAE candidates as a function of \lya line flux, and provide suggestions for color-magnitude selection cuts using medium-band photometry.
\end{itemize}

\begin{figure*}
    \centering
    \includegraphics[width=\linewidth]{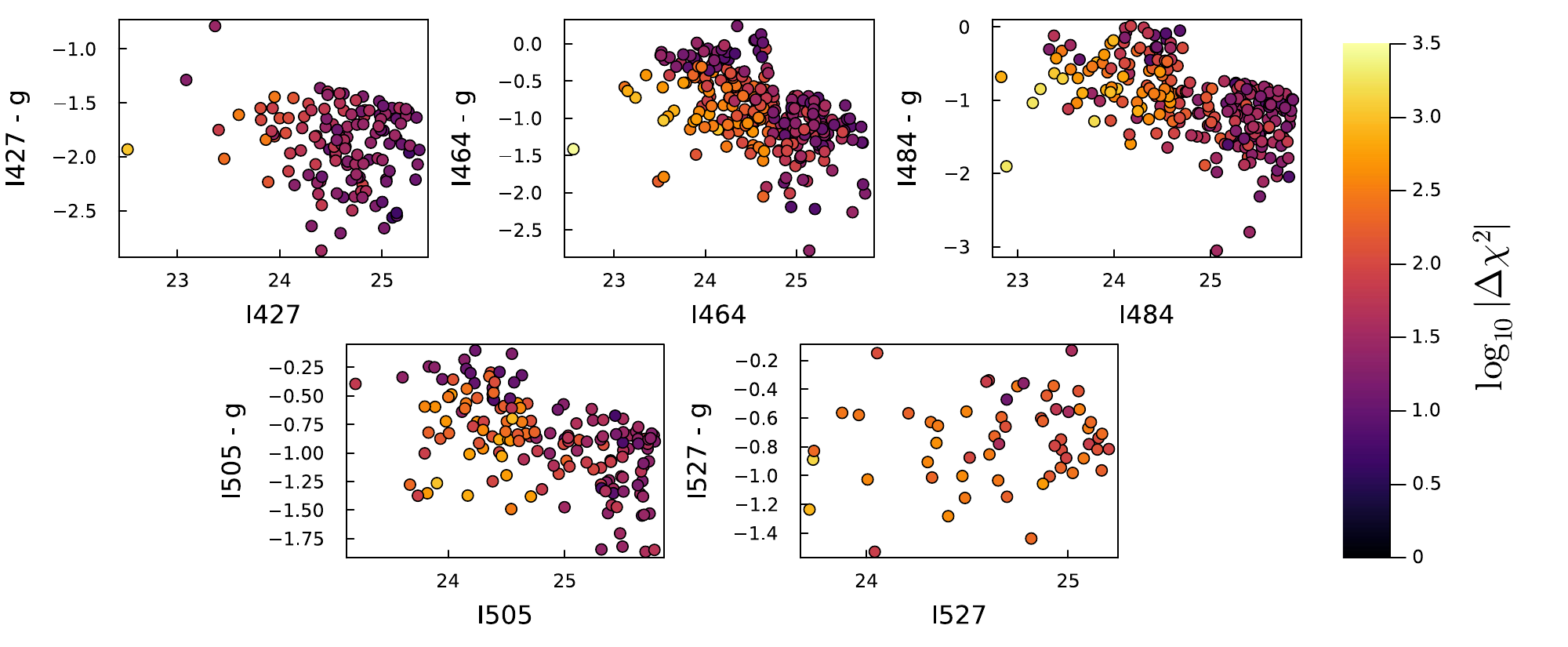}
    \caption{Color-magnitude diagrams for five Subaru Suprime-Cam medium bands (I427, I464, I484, I505, and I527) for \textbf{881} Subaru LAE targets, colored by $\log_{10}| \Delta \chi^2|$. Lighter point colors indicate higher $|\Delta \chi^2|$ and thus more confidence that the target is an LAE.}
    \label{fig:targeting}
\end{figure*}

We now present a framework to estimate the effective exposure time needed to constrain the redshift of an LAE candidate. DESI targets are observed with some number of variable-length exposures (aiming for uniform SNR in each exposure). These individually extracted spectra are then coadded using a per-wavelength inverse-variance weighted mean. To test the effect of shorter aggregate exposure time on the quality of our redshift estimates, we can coadd subsets of DESI exposures to simulate various realizations of a shorter exposure time for a given target. 

We first select a subset of all targets that were observed with 15-24 exposures and did not contain ``LBG" in the VI comments, leaving \textbf{310} targets. Each target then has a set of at least 15 individual exposure spectra, their corresponding inverse variance values, and the length of each exposure. For $n = 1\dots15$ exposures, we bootstrap (with replacement) $n$ exposures for each object, coadd them using inverse-variance weighting, and determine the redshift using our pipeline, using the \textbf{inverted} sum of the inverse variances on the diagonal of the  residual covariance. For each $n$, we calculate the standard deviation $\sigma_z$ of the redshift estimates and the average total exposure time (added across each subset of exposures) across 10 random sets of $n$ exposures.

Figure~\ref{fig:exposures} shows the cumulative fraction of these bootstrapped observations with converged redshift estimates ($\sigma_z < 0.001$) within bins of average exposure time and \lya line flux \citep{fastspecfit}. Targets with higher \lya line flux are more likely to have converged redshift estimates even with less than one hour of exposure time. As would be expected, objects with weaker \lya line flux need longer exposure times to achieve a stable redshift estimate. For some targets with line flux $< 10^{-16}~{\rm erg\, s^{-1}\, cm^{-2}}$, even combining all 15 exposures to yield an effective exposure time of more than 3.5 hours did not yield a converged redshift estimate.

Figure~\ref{fig:targeting} shows color-magnitude diagrams for medium bands from the Subaru Suprime-Cam (shown in Figure~\ref{fig:filters}), indicating potential targeting cuts that could be made to find additional LAE candidates using MB photometry (see \citet{white_clustering_2024} for previous LAE selection cuts). Objects with smaller medium band excess and fainter broadband magnitude consistently have lower $|\Delta \chi^2|$ values, indicating less confident LAE fits. Additional information about their magnitude could indicate if they are simply faint LAEs that require more exposure time, or if they are not LAEs at all. Potential cuts in color-magnitude space to yield a high-purity sample of LAEs could be MB mag $ < $ 24 and (MB-g mag) $<-1$, but more work should be done to refine this recommendation for specific filter combinations.

\section{Discussion}\label{sec:discussion}

\subsection{Pros and cons of this analysis}

Our data-driven prior covariances for the sky residual and LAE components provide a number of advantages. Explicitly modeling every potential sky residual line or the detailed profile of the \lya emission line would likely be difficult and time-consuming. Using data-driven priors for our sky residual component allows for easy capture of these systematic effects, which improves the performance of our method. By fitting a joint posterior and extracting the mean LAE component, we effectively marginalize over the sky residual component instead of subtracting out a point estimate. This presents a potential improvement over traditional sky subtraction methods, which can leave residual lines. 

Figure~\ref{fig:lae_eigenvecs} shows that we are able to express many physical nuances of the \lya line, such as the shape and asymmetry of the line profile. Each eigenvector appears to capture a different aspect of the line, including derivative shapes and adjustments to the amplitude and width. Every target spectrum has an associated set of coefficients for each eigenvector, which could be correlated with observed properties to provide additional insight into the physical nature of LAEs and their environments.

By evaluating Equation~\ref{eq:deltachisq} for many test redshifts, our pipeline yields not only a redshift estimate, but a $\Delta \chi^2$ profile as a function of redshift. In addition to reporting the redshift that minimizes $\Delta \chi^2$, we can use the best $|\Delta \chi^2|$ as a proxy for the SNR of the detection. To provide even more robust redshift predictions, the second- or third-best redshift values (that yield the second- and third-lowest $\Delta \chi^2$ values) could also be reported to provide alternate options. These $\Delta \chi^2$ values can be compared to that of the best-fit redshift. If they are very similar, it indicates an ambiguous fit and potential reason for skepticism of the redshift estimate. This mirrors a similar calculation done by the automated redshift determination pipeline from DESI used on the four main targeted classes to determine confidence in the redshift estimate \citep{Guy+23}.

\subsection{Pros and cons of MADGICS}

One of the main limitations of our method is that its performance relies heavily on the fidelity of the component prior covariance matrices, i.e., that the user knows the components they are looking for. It is relatively straightforward to derive high quality priors if high-SNR data for each component is available (see Equation~\ref{eq:covmatrix}). In practice, the component priors are carefully created over many iterations to ensure a satisfactory decomposition. Component priors can also be made using simulated data or theoretical models, or a hybrid of real and modeled data. However, any discrepancies between theory and data will be replicated when applying these priors to real data.

Despite its reliance on priors, MADGICS can still be applied to discover new data features or anomalies. Given that all components sum exactly to the data, anything that is not explicitly included as a component will fall into the residual component. These residual components can then be inspected or clustered to identify trends or anomalies. Additionally, if the gap between theory and data can be expressed as a linear combination, an additional component could be fit for that purpose. For an exploratory use case where the specific nature of the components in the data are totally unknown, a blind approach such as PCA or ICA may be a better choice.

MADGICS makes the fundamental assumptions that the data is a linear combination of mostly known components, and that the prior on each component can be expressed solely by a mean vector (which in practice is often set to zero) and covariance matrix. Nonlinearities can be expressed \citep[see][]{Saydjari+23}, but only via linear perturbation around an approximate solution. The assumption of Gaussianity within components is a safe one for spectroscopic data with thousands of wavelength bins, assuming that the number of bins is far greater than the number of relevant eigenvectors.

As with any numerical method, the derivation of uncertainties is critical to understanding its performance. In Sections~\ref{sec:uncertainty} and \ref{sec:simulations} we discuss how we determine redshift uncertainties based on the curvature of the \dchisq{} surface and validate them using injection tests. These uncertainties assume that the pipeline has determined the correct redshift and are not informative in the case of catastrophic failures. Additionally the MADGICS formalism encompasses not only posterior means for each component but also posterior covariances, which can be calculated as
\begin{equation}
    \hat{C}_{ii} = C_i - C_iC_{\rm tot}^{-1} C_i
\end{equation}
for component $i$ with prior covariance $C_i$ \citep{Saydjari+23}. The ability to determine a full posterior distribution over each component allows for more significant uncertainty quantification and robustness of results. This provides an advantage in particular over neural-network based methods, where it may be unclear how to derive a redshift uncertainty and ensure that it is properly calibrated.

\subsection{Future Work}

LAEs are one of the more difficult use cases for this method, because they only have one signature emission line instead of a more recognizable spectral feature or shape. The single \lya line can easily be confused with other emission lines by many methods, including ours; however, we rely that the \lya line shape recovered by our data-driven prior is unique enough to differentiate it from other emission lines. This could be further analyzed by running our pipeline on LAE spectra in the presence of more significant contaminants, such as quasars, [OII] emitters, and other emission-line galaxies using data-driven priors and levaraging their more complex correlations.

The performance of our method for LAEs suggests that it might be more successful in identifying other distinct spectral features, such as the Na doublet or Ca triplet. LBGs would be a natural next application of this framework, as there are already many identified LBGs that could be used to create a component prior, and they are also an anticipated priority for DESI-2. 

Our method can be applied to many astrophysical use cases outside of individual emission lines. MADGICS can fit multiple eigenvectors for each component, meaning it can encapsulate much more complex spectra, such as AGN spectra. A demonstration of the full expressivity of this method is forthcoming in future work.
 MADGICS can also be generalized for applications in photometry, with pixel-pixel covariance matrices for applications to star-galaxy separation and crowded-field photometry. 
 
Spectroscopy can be used to identify cases of strong gravitational lensing \citep[i.e.,][]{bolton_sloan_2006, huang2025desistronglensfoundry}, and our method could be generalized to automatically identify potential lensing candidates. To analyze LAE spectra, the number of components to fit is fixed, but the method can easily be extended to search for spectroscopic lenses by continuing to fit more emission line components (\lya or anything else for which we have a covariance prior) until the $\Delta \chi^2$ is reduced to the level expected for noise. Developing an automated procedure for spectroscopic lens identification would allow for scalable analysis of millions of spectra and likely additional detections of lensed systems. Assuming 0.4 seconds per source, it would take $\sim$6,000 CPU hours to process all 50 million DESI galaxy spectra, which could be spread across multiple cores to further reduce computational time.

\section{Conclusions} \label{sec:conclusion}

Here we present a new method for automated identification and spectroscopic redshift determination for LAEs in DESI, and apply it to provide insights into survey design in anticipation of DESI-2 and its emphasis on high-redshift targets. 

We use MADGICS, a Bayesian component separation method, to decompose each LAE target spectrum into sky residual, LAE, and residual components. We create a data-driven prior for the LAE component using spectra of LAE candidates targeted with narrow-band photometry from the ODIN survey. To determine a spectroscopic redshift for each target, we minimize the \dchisq{} from incorporating an LAE component while marginalizing over the sky residual lines. Our method is effective in automatically determining accurate spectroscopic redshifts for Subaru-targeted LAEs observed with DESI. Figure~\ref{fig:chisq} shows that we can determine redshift to within 0.005 of the visually inspected value for $>$90\% of targets. We additionally validate our method  by injecting a median LAE spectrum into both constant noise and real sky residual spectra, and provide a calibrated uncertainty for each redshift estimate. 

Our method has a number of useful applications to inform how future surveys can be optimized for observing LAEs. Our redshift pipeline returns a $|\Delta \chi^2|$ value that indicates the strength of the LAE detection, which can be used for targeting and catalog creation. By deriving redshifts for the same target by combining subsets of its exposures, we can assess how much exposure time is necessary for a given \lya line flux to ensure accurate LAE identification and characterization. We correlate the $|\Delta \chi^2|$ value derived from our method with color-magnitude diagrams from Subaru to make recommendations on potential targeting cuts using medium-band photometry ($\sim$250 \AA\ width, instead of using $\sim$100~\AA\ wide narrow-band filters) to optimize the redshift success rate when spectroscopically observing these LAE candidates. This is especially relevant in light of the upcoming DESI-2 mission and its planned focus on LAEs and other high-redshift objects.

Overall, our pipeline is robust for LAE identification and automated redshift determination, and is broadly generalizable to many additional applications inside and outside of astrophysical spectroscopy. We apply this method to explore potential implications for optimization of observational parameters and LAE targeting with medium-band photometry in anticipation of \mbox{DESI-2} and other large-scale high-redshift spectroscopic surveys.

\begin{acknowledgments}
\textbf{The authors thank the anonymous referee for valuable feedback.}

A.S.M.U. was supported by a National Science Foundation Graduate Research Fellowship and would like to thank Lisa Kewley, Joan Najita, Zihao Wu, Allyson Brodzeller, Aneta Siemiginowska, Daniel Eisenstein, and Cora Dvorkin for helpful discussions and suggestions.

A.K.S. acknowledges support by a National Science Foundation Graduate Research Fellowship (DGE-1745303) and that support for this work was provided by NASA through the NASA Hubble Fellowship grant HST-HF2-51564.001-A awarded by the Space Telescope Science Institute, which is operated by the Association of Universities for Research in Astronomy, Inc., for NASA, under contract NAS5-26555

A.D.’s research activities are supported by the
NSF NOIRLab, which is managed by the Association of
Universities for Research in Astronomy (AURA) under a
cooperative agreement with the National Science Foundation.

This work is supported by the National Science Foundation under Cooperative Agreement PHY-2019786 (The NSF AI Institute for Artificial Intelligence and Fundamental Interactions, \href{http://iaifi.org/}{http://iaifi.org/}).

This research used resources of the National Energy Research Scientific Computing Center (NERSC), a Department of Energy Office of Science User Facility.

This material is based upon work supported by the U.S. Department of Energy (DOE), Office of Science, Office of High-Energy Physics, under Contract No. DE–AC02–05CH11231, and by the National Energy Research Scientific Computing Center, a DOE Office of Science User Facility under the same contract. Additional support for DESI was provided by the U.S. National Science Foundation (NSF), Division of Astronomical Sciences under Contract No. AST-0950945 to the NSF’s National Optical-Infrared Astronomy Research Laboratory; the Science and Technology Facilities Council of the United Kingdom; the Gordon and Betty Moore Foundation; the Heising-Simons Foundation; the French Alternative Energies and Atomic Energy Commission (CEA); the National Council of Humanities, Science and Technology of Mexico (CONAHCYT); the Ministry of Science, Innovation and Universities of Spain (MICIU/AEI/10.13039/501100011033), and by the DESI Member Institutions: \url{https://www.desi.lbl.gov/collaborating-institutions}. Any opinions, findings, and conclusions or recommendations expressed in this material are those of the author(s) and do not necessarily reflect the views of the U. S. National Science Foundation, the U. S. Department of Energy, or any of the listed funding agencies.

The authors are honored to be permitted to conduct scientific research on I'oligam Du'ag (Kitt Peak), a mountain with particular significance to the Tohono O’odham Nation.

\end{acknowledgments}

%

\vspace{5mm}
\facilities{Mayall (DESI), Blanco (DECam)}


\software{Astropy \citep{astropy},  
          Julia \citep{bezanson2017julia}, 
          FastSpecFit \citep{fastspecfit}
          }

The data for all figures in this paper are available at \doi{10.5281/zenodo.15103552}.
\clearpage
\begin{appendix}
\section{Woodbury Updates for Computational Speedup}

Often, the limiting factor for covariance-based methods such as ours is the \dchisq{} calculation, which can be computationally expensive due to the matrix inversion. Here we calculate the $\chi^2$ as
\begin{equation}
\chi^2 = D^T C_{\rm tot}^{-1} D
\end{equation}

 We can rewrite the \dchisq{} calculation using matrix identities to avoid ever explicitly inverting a dense matrix. \cite{woodbury} presents the Woodbury matrix identity as follows:
\begin{equation}
(A + UCV)^{-1} = A^{-1} - A^{-1}U(C^{-1} + VA^{-1}U)^{-1}VA^{-1}
\end{equation}
This can be used to write the inverse of a matrix $A$ being modified by a low-rank matrix expressed by $VV^T$ (setting $U = V^T$ and $C = \mathbb{I})$:
\begin{equation}
(A + V V^T)^{-1} = A^{-1} - A^{-1}V(\mathbb{I} + V^TA^{-1}V)^{-1}V^TA^{-1}
\end{equation}
where $A$ is $n \times n$ and $V$ is $n \times k$, where $k$ is the number of eigenvectors represented by $V$. 

For our application of MADGICS to LAEs, $C_{\rm sky}$ and $C_{\rm LAE}$ are the same for every target for a given redshift, while $C_{\rm res}$ has the target-specific variances along the diagonal (see Section 3.2). 
In the 2-component case, $C_{\rm tot} = C_{\rm res} + V V^T$, and we can rewrite the $\chi^2$ calculation in Equation~\ref{eq:chisq} for an $n \times p$ data matrix D as:
\begin{align}
    \chi^2 &= D^T[(C_{\rm res} + VV^T)^{-1}] D \\  
    &= D^T[C_{\rm res}^{-1} - C_{\rm res}^{-1}V(\mathbb{I} + V^T C_{\rm res}^{-1} V)^{-1}V^TC_{\rm res}^{-1}] D
\end{align}
We seek to calculate the difference in $\chi^2$ from adding the LAE component, as in Equation \ref{eq:deltachisq}, which can be expressed as
\begin{align}
    \Delta \chi^2 &= -\left[(D^T \left(C_{\rm sky} + C_{\rm res}\right)^{-1} D - D^T \left(C_{\rm sky} + C_{\rm res} +  C_{\rm LAE}\right)^{-1} D\right]\\
    &= -D^T \left[ \left(C_{\rm sky} + C_{\rm res}\right)^{-1} - \left(C_{\rm sky} + C_{\rm res} +  C_{\rm LAE}\right)^{-1}\right] D
\end{align}
If we express $C_{\rm sky}$ and $C_{\rm LAE}$ as low-rank approximations
\begin{equation}
    C_{\rm sky} = V_{\rm sky} V_{\rm sky}^T
\end{equation}
\begin{equation}
    C_{\rm LAE} = V_{\rm LAE} V_{\rm LAE}^T
\end{equation}
and define the intermediary matrix
\begin{equation}
    M = C_{\rm res} + V_{\rm sky} V_{\rm sky}^T
\end{equation}
then we can rewrite \dchisq{} as
\begin{align}
    \Delta \chi^2 &= -D^T \left[ M^{-1} - \left(M +  V_{\rm LAE} V_{\rm LAE}\right)^{-1}\right] D \\
    &= -D^T \left[ M^{-1}V_{\rm LAE}(\mathbb{I} + V_{\rm LAE}^TM^{-1}V_{\rm LAE})^{-1}V_{\rm LAE}^TM^{-1}\right] D 
\end{align}
where $M^{-1}$ is then
\begin{equation}
    M^{-1} = C_{\rm res}^{-1} - \left[ C_{\rm res}^{-1}V_{\rm sky}(\mathbb{I} + V_{\rm sky}^TC_{\rm res}V_{\rm sky})^{-1}V_{\rm sky}^TC_{\rm res}^{-1}\right]
\end{equation}

In this way, we can calculate the \dchisq{} without ever inverting a dense matrix ($C_{\rm res}$ is diagonal), which provides significant computational speedup. 

\end{appendix}
\clearpage


\bibliography{main}{}
\bibliographystyle{aasjournal}



\end{document}